\documentclass{aa}
\usepackage{graphicx}
\usepackage[varg]{txfonts}
\usepackage{natbib,twoopt}
\usepackage[breaklinks=true,draft]{hyperref}
\usepackage{xcolor}
\usepackage{setspace}
\bibpunct{(}{)}{;}{a}{}{,}

\begin{document}

\title{A multi-epoch spectroscopic study of the BAL quasar\\APM 08279+5255}
\subtitle{II. Emission- and absorption-line variability time lags}

\author{F.~G. Saturni\inst{1}\fnmsep\inst{2}
\and
D. Trevese\inst{1}
\and
F. Vagnetti\inst{3}
\and
M. Perna\inst{4}
\and
M. Dadina\inst{5}}
\institute{Universit\`a degli Studi di Roma ``La Sapienza'', p.le A. Moro 5, I-00185 Rome (Italy)\\
\email{saturnfg@roma1.infn.it}
\and
Universit\`a degli Studi ``Roma Tre'', via della Vasca Navale 84, I-00146 Rome (Italy)
\and
Universit\`a di Roma ``Tor Vergata'', via della Ricerca Scientifica 1, I-00133 Rome (Italy)
\and
Universit\`a di Bologna, via Zamboni 33, I-40126 Bologna (Italy)
\and
INAF-IASF Bologna, via Gobetti 101, I-40129 Bologna (Italy)}

\date{Received 2015 August 10; accepted 2015 December 4}

\abstract
   {The study of high-redshift bright quasars is crucial to gather information about the history of galaxy assembly and evolution. Variability analyses can provide useful data on the physics of the quasar processes and their relation with the host galaxy.}
   {In this study, we aim at measuring the black hole mass of the bright lensed BAL QSO APM 08279+5255 at $z = 3.911$ through reverberation mapping, and at updating and extending the monitoring of its  C {\tiny IV} absorption line variability.}
   {Thanks to 138 $R$-band photometric data and 30 spectra available over 16 years of observations, we perform the first reverberation mapping of the Si {\tiny IV} and C {\tiny IV} emission lines for a high-luminosity quasar at high redshift. We also cross-correlate the C {\tiny IV} absorption equivalent width variations with the continuum light curve, in order to estimate the recombination time lags of the various absorbers and infer the physical conditions of the ionised gas.}
   {We find a reverberation-mapping time lag of $\sim 900$ rest-frame days for both Si {\tiny IV} and C {\tiny IV} emission lines. This is consistent with an extension of the BLR size-to-luminosity relation for active galactic nuclei up to a luminosity of $\sim 10^{48}$ erg s$^{-1}$, and implies a black hole mass of $10^{10}$ $M_\odot$. Additionally, we measure a recombination time lag of $\sim 160$ days in the rest frame for the C {\tiny IV} narrow absorption system, which implies an electron density of the absorbing gas of $\sim 2.5 \cdot 10^4$ cm$^{-3}$.}
   {The measured black hole mass of APM 08279+5255 indicates that the quasar resides in an under-massive host-galaxy bulge with $M_{bulge} \sim 7.5 M_{BH}$, and that the lens magnification is lower than $\sim 8$. Finally, the inferred electron density of the narrow-line absorber implies a distance of the order of 10 kpc of the absorbing gas from the quasar, placing it within the host galaxy.}


\titlerunning{Variability time scales in APM 08279+5255}
\authorrunning{F.~G. Saturni et al.}
\maketitle

\section{Introduction}\label{intro}

Quasars are broad-line, high-luminosity active galactic nuclei (AGNs) with spectral features surprisingly similar to those of the Seyfert galaxies, despite a difference in luminosity of several orders of magnitude. This implies a common process at the base of their existence. Moreover, quasars distinguish themselves for emitting at all wavelengths from radio to gamma rays, with a total energy emission comparable to that of a bright galaxy. The commonly accepted model able to match the observations is the accretion of gas around a supermassive black hole (SMBH; \citealt{Sal64,Zel65}).

The similarity in quasar spectra naturally leads to build a unified scenario for the structure of quasars (see e.g. \citealt{Urr95} for a review). In the simplest one, ionised gas clouds orbit around the black hole at radii larger than the accretion disk responsible of the continuum emission, producing the broad (from a broad-line region, BLR, still relatively close to the central black hole) and narrow (from a farther narrow-line region, NLR) emission lines. At intermediate orientations between the accretion disk plane and its axis, high-velocity winds can be launched under the action of the radiation pressure, thus originating the broad-absorption line (BAL) phenomenon when the quasar is viewed along the wind \citep{Wey83,Tur88,Elv00}.

Assuming that such features are common to all quasars, the only two parameters which the quasar structure depends on are the SMBH mass and the quasar inclination angle with respect to the line of sight. The first quantity has great relevance in cosmological studies: in fact, SMBHs trace large-scale structures and co-evolve with their host galaxies (see e.g. \citealt{Cat09} and refs. therein), thus giving information about the quantity of matter in the Universe and the history of galaxy assembly.

Currently, the only direct way to estimate the SMBH mass beyond the local Universe is through reverberation mapping (RM; \citealt{Gas87,Ede88,Whi94,Pet97}), by which the time lag $t_{lag}$ between the quasar broad emission-line and continuum variability, hence the size of the BLR $R_{BLR} = ct_{lag}$, can be measured. Under the hypothesis of virialised orbits in the BLR, the black hole mass $M_{BH}$ is given by:

\begin{equation}\label{revmapeq}
M_{BH} = f \frac{ct_{lag} \Delta v^2}{G}
\end{equation}

Here, the form factor $f$ summarises all the ignorance about the shape and inclination with respect to the line of sight of the BLR, that can significantly alter the final value of $M_{BH}$, while $\Delta v$ is the full width at half maximum (FWHM) of the considered emission line. Under reasonable assumptions on the distribution of gas clouds in the BLR (i.e., assuming some physical value for $f$), this formula gives a measurement of $M_{BH}$ in principle for all classes of active galactic nuclei (AGNs) with broad emission lines. In this way, relations between mass, line width and quasar luminosity (e.g., \citealt{McL04,Ves06,She11}) can be computed, and the SMBH mass can be estimated from single-epoch quasar spectra. This avoids the necessity of a RM for each object, allowing to obtain mass estimates for a large number of sources.

The application of the RM method was initially limited to low-luminosity, local AGNs like the Seyfert galaxies, that have variability time scales from days to weeks at most \citep{Wan99,Pet04}. Subsequently, it was applied to low-redshift quasars ($z < 0.4$; \citealt{Kas00}). For higher-luminosity, higher-redshift quasars, the increase of the BLR size with luminosity and the cosmological time dilation increase the variability time scales up to years, thus making campaigns for quasar RM observationally expensive. Until now, only three luminous, high-redshift quasars have RM mass measurements \citep{Kas07,Che12,Tre14}; this lack of information at high masses and luminosities means that the application of single-epoch mass-luminosity relations to quasars is actually an extrapolation based on data from AGNs of moderate redshift only, making subject to possible biases the current black hole mass estimates of quasars in cosmological surveys (e.g., \citealt{She11}).

Broad absorption-line quasars (BAL QSOs; \citealt{Lyn67,Wey83,Tur88,Elv00}) are a very important test for unified scenarios, since the mechanisms to launch a highly ionised, high-velocity (up to $\sim 0.2 c$) absorbing outflow are still poorly known. The exact location, structure and physical properties of BAL outflows  are difficult to introduce in unification schemes for quasars.

BAL variability (e.g., \citealt{Bar92,Gib08,Cap11,Tre13}) is one of the main ways to gather information about the gas producing the broad absorption features, such as physical (density, temperature, ionisation) and geometrical (location, opening angle, bending) properties of the outflow. Several programmes to study BAL variability are going on, based either on single objects with multi-year observations \citep{Bar92,Kro10,Hal11,Tre13,Gri15} or on {\itshape ensemble} analyses of quasar samples with a few (up to 10) observations per object \citep{Bar93,Lun07,Gib08,Cap11,Cap12,Cap13,Fil13}. Currently, the possible explanation of the BAL variability over multi-year time scales consider changes either in the gas ionisation level or in the covering factor, although the second possibility is favoured since: (i) the variations tend to occur in narrow portions of BAL troughs (e.g., \citealt{Fil13}), and (ii) they generally do not correlate with changes in the observed continuum (\citealt{Gib08,Cap11}; but see \citealt{Tre13}).

In this study, we take advantage of our long-term monitoring programme of high-luminosity quasars \citep{Tre07} to increase the data statistics about the bright BAL QSO APM 08279+5255, in order to update our study on its C {\scriptsize IV} absorption variability (\citealt{Tre13}, Paper I henceforth; \citealt{Sat14}) and to obtain estimates of its black hole mass through RM of Si {\scriptsize IV} and C {\scriptsize IV} emission lines. APM 08279+5255 is one of the most luminous quasars, discovered in 1998 by \citet{Irw98}. It is well known for its gravitational lensing, first confirmed case of a lens with odd number of image components \citep{Lew02}. Additionally, its redshift inferred from the high-ionisation emission lines (N {\scriptsize V}, Si {\scriptsize IV}, C {\scriptsize IV} and C {\scriptsize III}]) is $z =3.87$ \citep{Irw98}, which corresponds to an outflow velocity of $\sim 2500$ km s$^{-1}$ with respect to the systemic redshift $z = 3.911$ derived from the CO emission lines \citep{Dow99}.

The paper is organised as follows: in Sect. \ref{odr}, we briefly present the target of the observations and describe the adopted observational strategy and data reduction; in Sect. \ref{rmap}, we present the line-to-continuum RM time lags; in Sect. \ref{avar}, we update the absorption variability study of APM 08279+5255 with new data; finally, in Sect. \ref{disc} we discuss APM 08279+5255 black hole mass estimates and absorption variability, summarising all the findings in Sect. \ref{conc}. Throughout this paper, we use the terms ``quasar'' and ``QSO'' interchangeably, and adopt a concordance cosmology with $H_0 = 70$ km s$^{-1}$ Mpc$^{-1}$, $\Omega_M = 0.3$ and $\Omega_\Lambda = 0.7$.

\begin{figure*}[htbp]
\begin{center}
\includegraphics[scale=0.7]{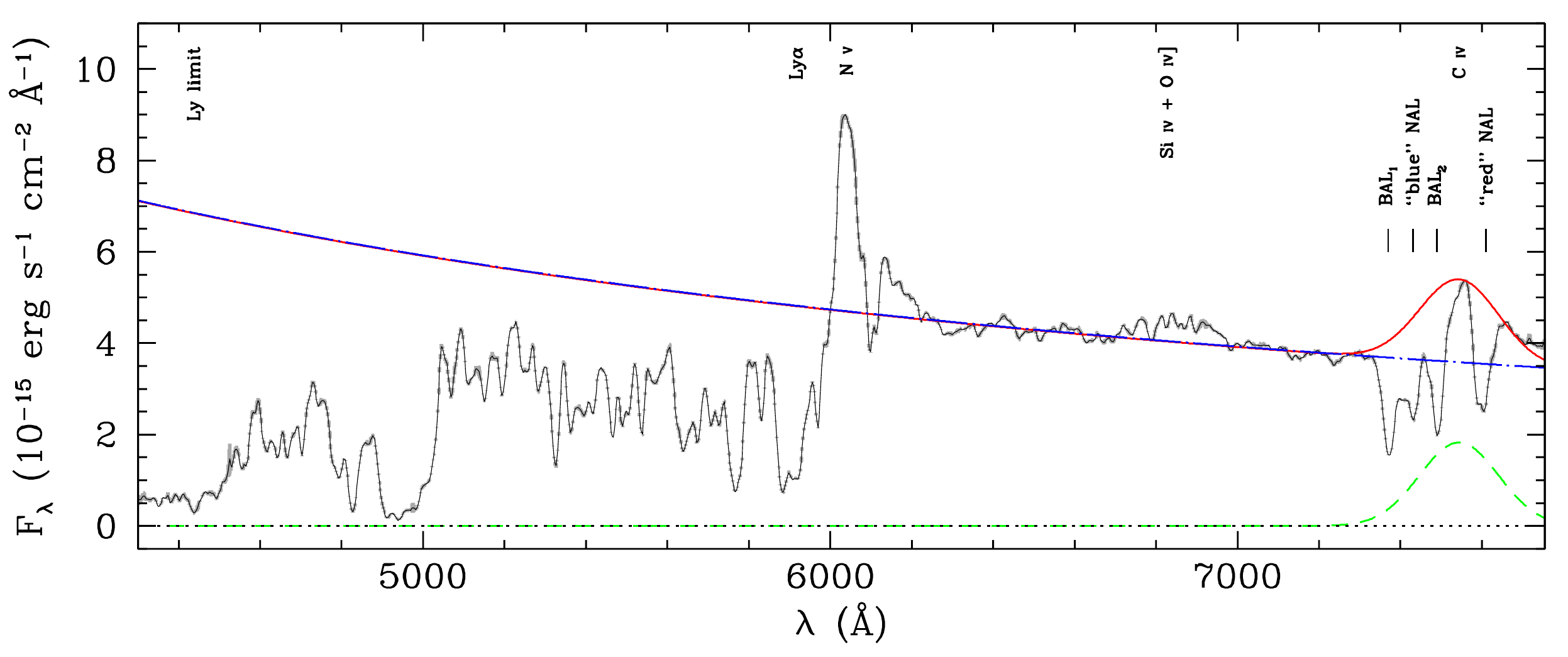}
\caption{The de-reddened spectrum of APM 08279+5255 taken at MJD = 52,695 at the 1.82 m Copernico telescope (Asiago, Italy). Superimposed to the spectrum ({\itshape black solid line}), a power-law continuum ({\itshape blue dot-dashed line}), a Gaussian fit of the C {\tiny IV} emission line ({\itshape green dashed line}) and the combination of these two components into a single pseudo-continuum ({\itshape red solid line}) are shown. The dotted line marks the flux zero level. The major absorption features are marked with the identifiers used in the text (see Sect. \ref{avar}) and ticks. The positions of the Lyman limit at $\lambda$912 \AA~in the rest frame and of the major emission lines are also indicated.}
\label{apmspec}
\end{center}
\end{figure*}

\section{Observations and data reduction}\label{odr}

\subsection{The reverberation mapping campaign}

Our RM campaign of luminous quasars started in 2003 at the Asiago Observatory, Italy \citep{Tre07}. Motivated by the work of \citet{Net03} about the possible biases introduced in galaxy mass estimates by an overestimation of their black hole mass $M_{BH}$, the goal was to measure $M_{BH}$ of some intrinsically bright objects in order to extend the BLR size-luminosity relationships for AGNs \citep{Pet05,Kas07} to luminosities greater than $\sim 10^{46}$ erg s$^{-1}$. Both $R$-band photometric data and spectra have been obtained with the Asiago Faint Object Spectrograph and Camera (AFOSC) at the 1.82 m Copernico telescope (Asiago, Italy). Since December 2012, the campaign is going on at the 1.52 m Cassini telescope (Loiano, Italy), with the Bologna Faint Object Spectrograph and Camera (BFOSC).

\begin{figure}[htbp]
\begin{center}
\includegraphics[scale=0.45]{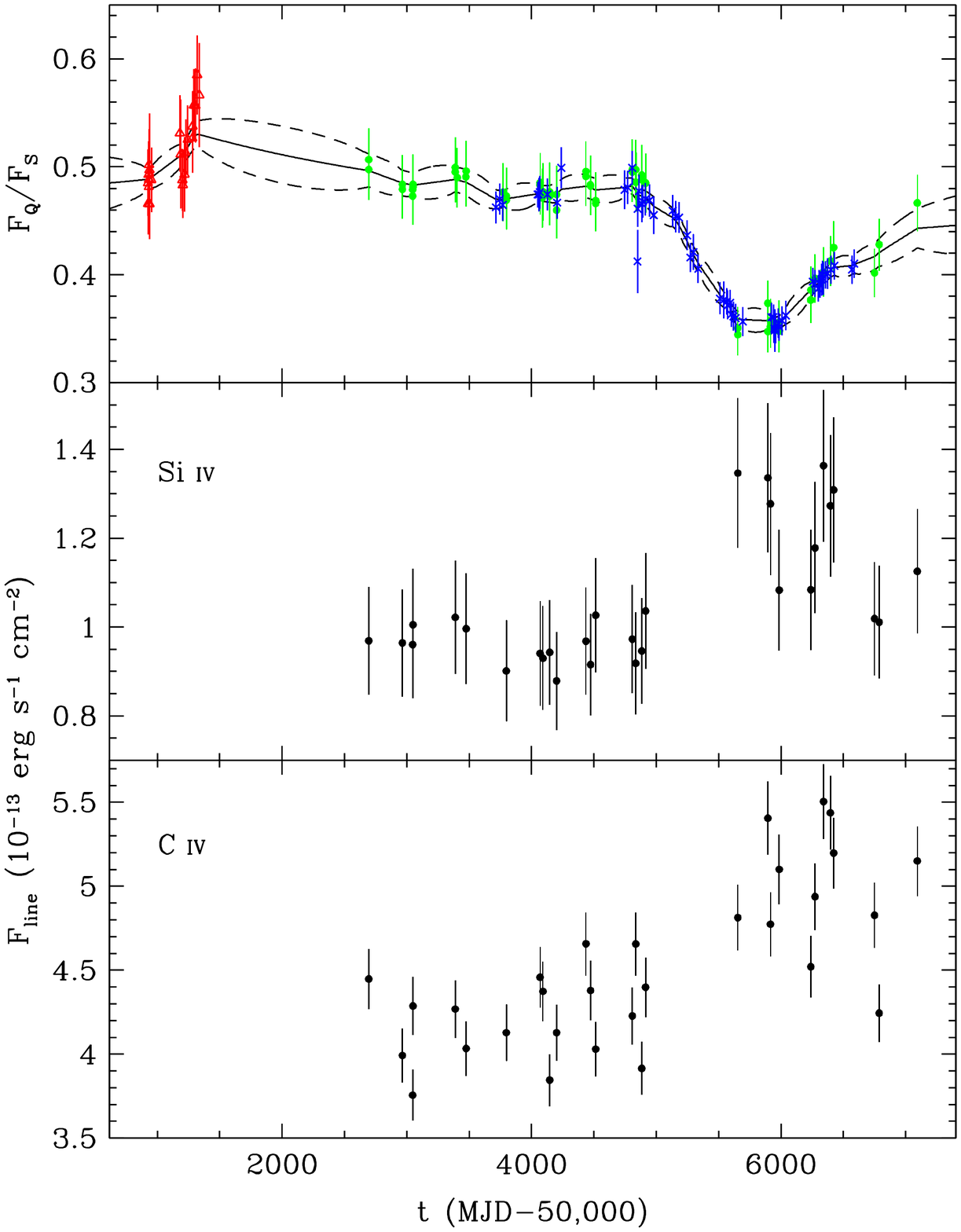}
\end{center}
\caption{{\itshape Upper panel:} $R$-band continuum light curve of APM 08279+5255. The \citep{Lew99} photometric campaign ({\itshape red triangles}), the Asiago and Loiano \citep{Tre07,Tre13,Tre14} RM campaign ({\itshape green dots}) and the rescaled V-band Catalina photometry ({\itshape blue crosses}) are all included. Superimposed to the data, the best-fit DRW variability model computed by JAVELIN \citep{Zu11} is shown ({\itshape black solid line}) delimited by its error boundaries at 68\% probability ({\itshape black dashed lines}). {\itshape Middle panel:} Si {\tiny IV} light curve. {\itshape Lower panel:} C {\tiny IV} light curve.}
\label{ligcur}
\end{figure}

The procedure of observation and data reduction is explained in detail in previous papers \citep{Tre07,Tre13,Tre14}. Here we simply recall that the instrument was set in order to simultaneously observe the quasar and a star $S$ of comparable magnitude ($R =14.66$; \citealt{Ros10}), located at $\alpha$ 08 31 22.3 $\delta$ +52 44 58.6 (J2000), which has been adopted as calibration reference for both photometry and spectra. The QSO and the star are observed in the same slit. A wide slit of 8'' (on AFOSC) or 12'' (on BFOSC) was adopted to avoid light losses due to small misalignments or differential refraction, that would produce artificial variability. In this way, we obtain differential magnitudes $\Delta R$ between the quasar and the reference star from the photometry, and spectral ratios $Q/S$ between the uncalibrated quasar ($Q$) and stellar ($S$) spectra. The latter are then multiplied to a stellar spectrum of $S$ taken at MJD = 55,894.5 and calibrated through standard IRAF techniques: assuming that the reference star is non-variable, this provides a flux calibration preserving the quasar variability. An example of these spectra is shown in Fig. \ref{apmspec}. In total, we have collected 28 $R$-band photometric data and 30 spectra until March 2015, part of which has been already analysed in previous papers focussed on the C {\scriptsize IV} absorption variability (Paper I; \citealt{Sat14}).

The RM technique requires to construct a quasar continuum light curve (LC) and an emission-line LC for each line considered. Therefore, we have converted our magnitude differences $\Delta R$ between quasar and reference star to flux ratios $F_Q/F_S$; since we compute an emission-line contribution to the total $R$-band flux of $\sim 11$\% only, we assume that our photometry is well tracing the quasar UV rest-frame continuum variability, and adopt it as the reference to construct APM 08279+5255 continuum LC. We have then computed the quasar continuum flux from our series of spectra in a rest-frame spectral interval of 100 \AA~around $\lambda = 1350$ \AA. The LC obtained in this way differs from the $R$-band photometry by only a scale factor; this, assumed constant for all spectra, has been computed over a time interval in which the quasar flux level remains approximately constant, and has been verified to provide a good match to the $R$-band photometry also in the presence of flux variations. Therefore, we have scaled by this factor the spectral continuum LC to the photometric data, and have added it to the full data set of the continuum LC. This improves our statistics to 58 points tracing the continuum variations of APM 08279+5255. In Tab. \ref{tabflux}, we report our spectro-photometric data together with the data collected from the literature described in the next subsection.

\subsection{Data from the literature}

As in Paper I, we have added the photometric data of \citet{Lew99} to our data set, after re-scaling their measurements from their reference star $S_1$, which is always serendipitously included in our field of view, to our reference star $S$. This provides us with additional 23 photometric points. One last data set of 59 $V$-band points, scaled to our $R$-band photometry in the same way of our spectral continuum fluxes, came from the Catalina Sky Survey \citep{Dra09}. In this way, we end up with a total of 138 continuum data points for APM 08279+5255, that along with 30 spectra constitute so far the largest data set collected for a quasar at such high redshift. Tab. \ref{tabflux} summarises our observations and the literature data: we give the observation date MJD $- 50,000$ in Col. 1, the telescope in Col. 2, the observation type (spectroscopic or photometric) in Col. 3, the continuum flux ratios $F_Q/F_S$ referred to the $R$-band fluxes (as explained in the previous subsection) in Col. 4, the Si {\scriptsize IV} and C {\scriptsize IV} emission-line fluxes in Col. 5 and 6 respectively (see Sect. \ref{rmap}).

We also take advantage of the existence of further five spectra of APM 08279+5255 available in the literature (\citealt{Irw98,Ell99,Hin99,Lew02}; Saturni et al. 2015, in prep.), introducing them into our monitoring as well. Unfortunately, due to the lack of a uniform flux calibration these spectra cannot be used in our RM, and hence will be added to the current data set only for the purposes of updating the C {\scriptsize IV} absorption variability study presented in Paper I and \citet{Sat14}, which requires only fluxes normalised to the local continuum. Only the HST/STIS spectrum from \citet{Lew02} has been used in the RM, but for the sole purpose of determining the C {\scriptsize IV} emission line width to be used in the RM measurement of APM 08279+5255 SMBH mass; this is motivated by the fact that this spectrum is taken outside the atmosphere, thus avoiding the telluric absorption by the Fraunhofer A band on the residual C {\scriptsize IV} emission-line wing.

\section{Reverberation mapping of the Si {\normalsize IV} and C {\normalsize IV} emission lines}\label{rmap}

Thanks to our long-time monitoring, we are now able to perform the RM of APM 08279+5255. This is done by cross-correlating the variability of the $R$-band flux tracing the continuum with the Si {\scriptsize IV} and C {\scriptsize IV} emission-line fluxes, in order to find the most probable lag between continuum and line variations. The LCs to be used in this process, shown in Fig. \ref{ligcur}, are constructed as follows:

\begin{itemize}

\item APM 08279+5255 spectrum is reddened by a foreground absorber at $z \sim 1$ (likely the so-far invisible lensing galaxy; \citealt{Ell04}), with a $V$-band reddening parameter $A_V \sim 0.5$ mag \citep{Pet00}: in order to obtain a better power-law fit to the continuum, we de-redden our series of spectra with a Small-Magellanic-Cloud-like extinction \citep{Pei92} produced at $z = 1.062$ \citep{Ell04}. This function is fixed at all epochs, hence it does not introduce any noise in the variability;

\item for all the spectra, we evaluate the Si {\scriptsize IV} flux by subtracting a power-law continuum computed at each epoch over two wavelength intervals, $\lambda\lambda 6620 - 6680$ and $\lambda\lambda 7000 - 7100$ in the observer frame, adjacent to that emission line and free of other emission or absorption features. The Si {\scriptsize IV} flux is then defined as the integral of the line flux in the interval $\lambda\lambda 6680 - 7000$;

\begin{figure}[htbp]
\begin{center}
\includegraphics[scale=0.45]{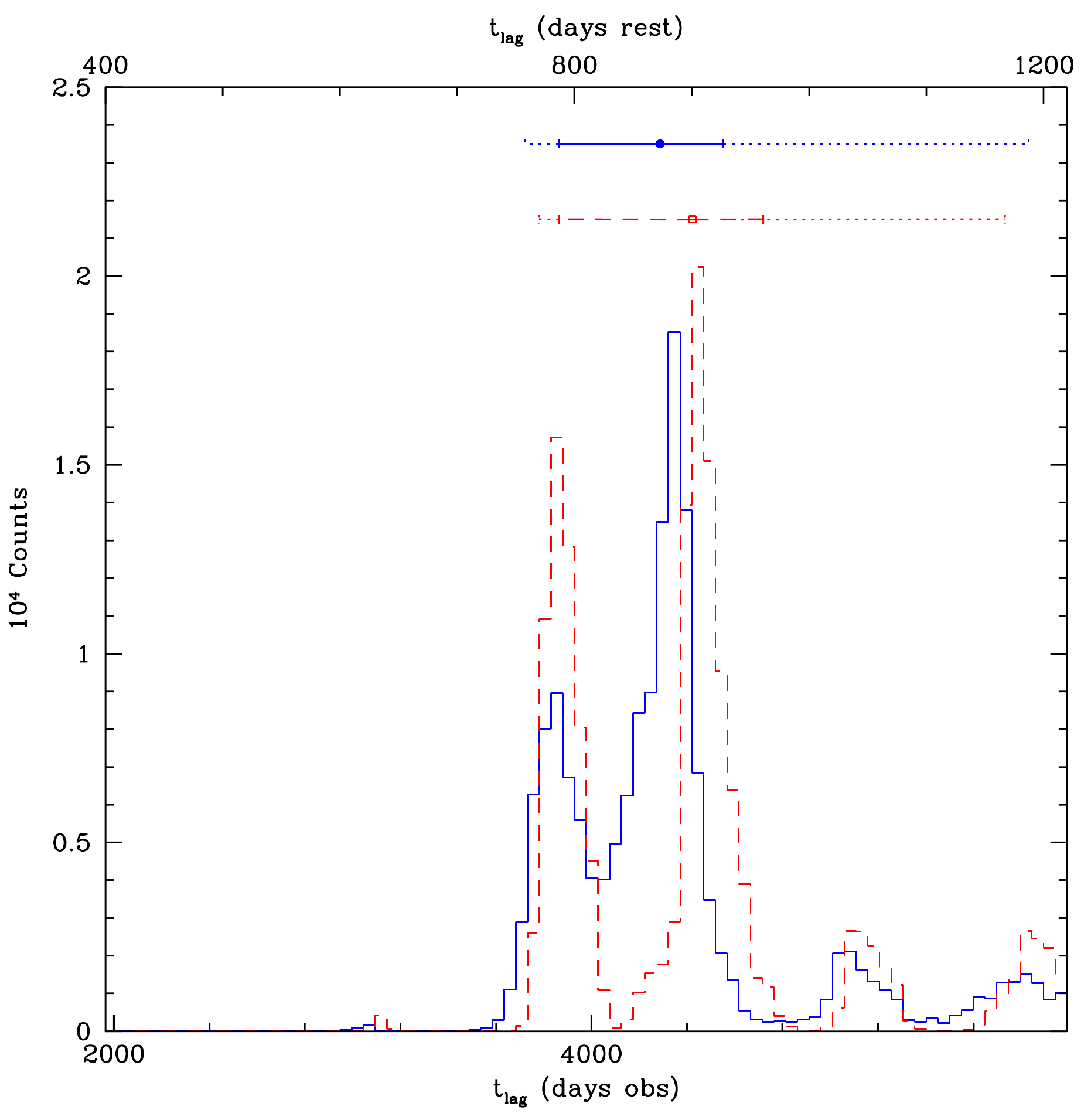}
\end{center}
\caption{Posterior distributions of the Si {\tiny IV} ({\itshape blue solid histogram}) and C {\tiny IV} ({\itshape red dashed histogram}) RM time lags computed with JAVELIN. The time-lag mean values ({\itshape blue point and red square}) for each distribution are shown, together with their corresponding confidence intervals at 68\% (colour and line-type codes as per the histograms) and at 95\% probability ({\itshape dotted error bars}).}
\label{laghpd}
\end{figure}

\item finally, we apply the same fitting procedure described in Paper I and \citet{Sat14} to the C {\scriptsize IV} emission line, assumed Gaussian, and compute its line flux as the integral of the resulting fitting function at each epoch. In this procedure, only the line amplitude is allowed to vary, whereas the peak position is fixed to the C {\scriptsize IV} $\lambda$1549 at $z = 3.87$, and the Gaussian width is fixed to the value $\sigma_g = 3200$ km s$^{-1}$ obtained through a fit performed onto the HST/STIS high-resolution spectrum \citep{Lew02}, which is free of the O$_2$ telluric absorption on the C {\scriptsize IV} red wing.
\end{itemize}

We report all the emission-line fluxes in the fifth and sixth columns of Tab. \ref{tabflux} together with their 1$\sigma$ errors.

In order to determine the emission-line time lag, we recall that the usual method consists in evaluating the peak or centroid of the line-to-continuum cross-correlation function (CCF). This can be constructed either by computing a discrete correlation function (DCF; \citealt{Ede88}) or via an interpolation of the LCs (ICCF; \citealt{Gas87,Whi94}). However, CCFs provide consistent results only for well-sampled LCs, instead presenting technical problems in the case of poor sampling \citep{Pet93,Wel99}. In particular, the DCF sensitivity to real correlation decreases. An estimate of the confidence interval on the measured time lag can be obtained through the DCF $z$-transformation method (ZDCF) by \citet{Ale97,Ale13}; however, this requires to eliminate from each DCF bin all points corresponding to pairs of epochs with measured data in common. Thus, this is hardly applicable in our case, where the sampling is particularly poor.

\begin{figure*}[htbp]
\begin{center}
\includegraphics[scale=0.7]{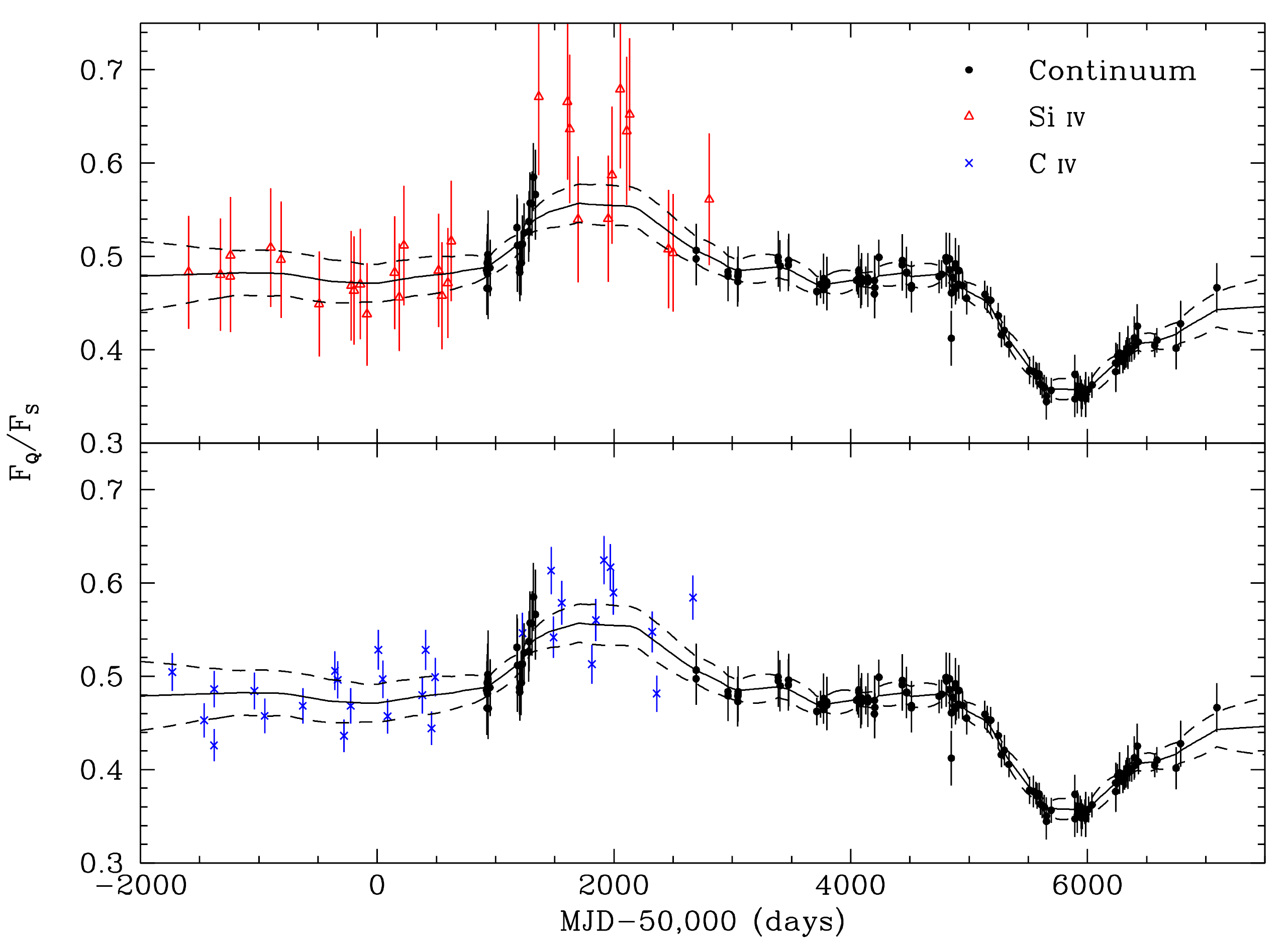}
\end{center}
\caption{APM 08279+5255 light curve composed with the continuum ({\itshape black dots}) and emission line fluxes, the latter scaled in intensity and shifted in time of their JAVELIN lags of 837 and 901 rest-frame days respectively to mimic the original continuum light curve that is reverberated by the quasar BLR. {\itshape Upper panel:} continuum and Si {\tiny IV} ({\itshape red triangles}). {\itshape Lower panel:} continuum and C {\tiny IV} ({\itshape blue crosses}). In both panels, a DRW continuum, fitted by JAVELIN to the data with the damping time scale $\tau_d$ blocked to that of the continuum fit, is shown superimposed to the light curve ({\itshape solid line}) along with its $1\sigma$ error band ({\itshape dashed lines}).}
\label{laglin}
\end{figure*}

We therefore evaluate the RM time lags of SI {\scriptsize IV} and C {\scriptsize IV} with the JAVELIN code\footnote{Available at {\ttfamily http://www.astronomy.ohio-state.edu/$\sim$yingzu\\/codes.html}.}, based on its previous version SPEAR (Stochastic Process Estimation for AGN Reverberation) by \citet{Zu11}. With respect to traditional CCFs, the advantage of this method, whose statistical bases date back to \citet{Pre92,Ryb92} with a first application in \citet{Ryb94}, is that it makes use of an interpolation algorithm in which the entire data set contributes to each interpolated point, under the assumption that the emission-line LCs $F_l(t)$ are a scaled, smoothed and delayed version of the continuum LC $F_c(t)$. The QSO continuum LC is then modeled with a damped random walk (DRW) process described by a damping time scale $\tau_d$ and an rms variability amplitude $\sigma$ \citep{Kel09,Koz10,Mac10,Zu13}. This is realised through the application of statistical weights that are determined from the auto-correlation functions of the data.

The emission-line LC is obtained from the continuum LC through the convolution with a transfer function $\Psi(t)$:

\begin{equation}\label{trans}
F_l(t) = \int \Psi(t')F_c(t-t') dt'
\end{equation} 
In this analysis, we adopt for $\Psi(t)$ the simple analytic top-hat form as done by \citet{Zu11}, of width $\Delta$ and area $A$:

\begin{equation}\label{tophat}
\Psi(t) = \frac{A}{\Delta},\mbox{ }\left|
t - t_{lag}
\right| \leq \Delta\mbox{, and zero elsewhere.}
\end{equation}
This choice is based on the fact that the resulting emission-line lag $t_{lag}$ does not depend strongly on the specific form of $\Psi(t)$ \citep{Ryb94}. A maximum-likelihood code provides the five best-fit parameters ($\tau_d$, $\sigma$, $t_{lag}$, $A$ and $\Delta$) for the DRW model that describes the continuum LC and the top-hat transfer function. Finally, confidence limits on the fit parameters are obtained through Monte-Carlo Markov chain (MCMC) iterations. With respect to traditional CCFs, the model-dependent nature of JAVELIN has the advantage of producing smaller lag uncertainties \citep{Zu11}.

\begin{figure}[htbp]
\begin{center}
\includegraphics[scale=0.45]{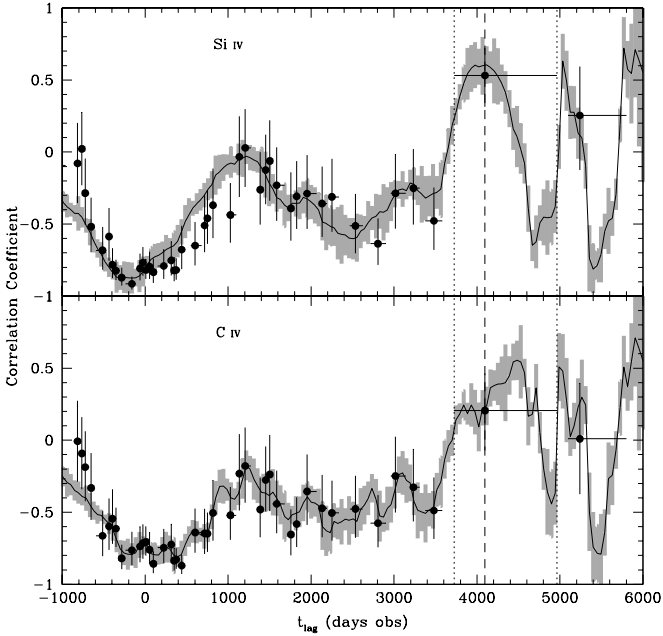}
\end{center}
\caption{{\itshape Upper panel:} ICCF ({\itshape solid line}) with its $1\sigma$ error ({\itshape grey area}) and ZDCF ({\itshape black points}) for the continuum and the Si {\tiny IV} emission line. {\itshape Lower panel:} same as in the upper panel, but for the continuum and the C {\tiny IV} emission line. The ZDCF is computed over 15 uncorrelated points per bin, and finds a simultaneous peak for the two emission features at $t_{lag} = 4095^{+867}_{-369}$ days in the observer frame that is marked in the plot ({\itshape dashed line} and {\itshape dotted lines}).}
\label{revmapccf}
\end{figure}

In order to verify that the result of JAVELIN does not depend on the assumed model, we compute the classical ICCF and DCF via its $z$-transformation (ZDCF) for each pair of line and continuum LCs, and evaluate the corresponding RM time lags with their uncertainties. Although in our case the CCFs have low statistical significance, they are completely model-independent.

The uncertainty on the ICCF lag is estimated by applying the ``flux randomisation and random sample selection'' method (FR/RSS) by \citet{Pet98}. This method allows to compute the statistical distribution of the line-to-continuum time lags by iterating the ICCF procedure over undersampled continuum and emission-line LCs, altering the remaining data by adding a random noise with a Gaussian distribution whose standard deviation is the 1$\sigma$ error on each measured point. In order to compute the ZDCF time lag and the uncertainty on the peak position, we use the routines made available online by T. Alexander\footnote{Both scripts are available at {\ttfamily http://wwo.weizmann.ac.il/\\weizsites/tal/research/software/}.}.

In Sect. \ref{single} and \ref{joint}, we respectively report the RM time lags found with the analysis done by JAVELIN, and those found through the CCFs for comparison.

\subsection{Reverberation mapping with JAVELIN}\label{single}

As a first step, we apply the JAVELIN method to study the correlation of the individual Si {\scriptsize IV} and C {\scriptsize IV} lines with the continuum. Due to our unprecedented long-time monitoring, we are able to probe time lags in the observer frame up to $\sim 6000$ days, corresponding to $\sim 1200$ rest-frame days at $z = 3.911$. The analysis with JAVELIN gives $t_{lag}^{JAV}\mbox{(Si {\scriptsize IV})} = 4180^{+1035}_{-323}$ days in the observer frame and $t_{lag}^{JAV}\mbox{(C {\scriptsize IV})} = 4243^{+337}_{-355}$ days, thus suggesting that the BLR emitting the Si {\scriptsize IV} and the C {\scriptsize IV} lines are both located at the same distance from the QSO central engine ($\sim 850$ light days for APM 08279+5255).

The RM time lags can be also estimated by fitting with JAVELIN the LCs of multiple emission lines together. This is particularly useful when the LC sampling is sparse, since, if all emission lines follow the same variability model as the continuum, the available information to fit the variability increases. Therefore, we perform this analysis with the full data set of 138 continuum points and 30 points for each emission line, for a total of $N_{tot} = 198$ light-curve points to fit the eight parameters (continuum damping time scale and variability amplitude, two top-hat lags, two smoothing widths and two scale factors) of the DRW plus top-hat transfer function model of quasar line-to-continuum variability.

The joint fit with JAVELIN gives $t_{lag}^{JAV}(\mbox{Si {\scriptsize IV}}) = 4286^{+260}_{-417}$ days in the observer frame and $t_{lag}^{JAV}(\mbox{C {\scriptsize IV}}) = 4425^{+280}_{-560}$ days respectively. In Fig. \ref{laghpd}, we show the posterior distribution of a JAVELIN MCMC run with $1.6 \cdot 10^5$ iterations for these time lags. Also, in Fig. \ref{laglin} we show the emission-line LCs of Si {\scriptsize IV} and C {\scriptsize IV} scaled to the $R$-band continuum, and shifted back in time of the relevant RM lags. At a first glance, the emission-line LCs are both consistent with the interpolated decrease in the continuum flux after the rise around MJD $\sim 51,500$ observed by \citet{Lew99}.

Some caution must be adopted in interpreting the result, since the lag associated to the cross-correlation peak corresponds to a maximum of the emission-line LCs just falling in a gap of the continuum LC (see Fig. \ref{laglin}). This is however made possible by the assumption of the method, which uses the information of the emission-line LCs to interpolate the continuum. In the continuum LC gap between 51,500 and 52,500 days, JAVELIN is therefore using the local maximum of the emission-line LCs at MJD $\sim 56,000$: although this is the probable cause of the extension at large time lags of the emission-line lag posterior distribution, and consequently of the asymmetric errors at 95\% probability in the lag estimate (see Fig. \ref{laghpd}), the result is quite plausible. With this measurement, APM 08279+5255 becomes the most distant object, and one of the intrinsically most luminous, for which RM time lags are available.

\subsection{Reverberation mapping with the CCFs}\label{joint}

For comparison with the time lags found by JAVELIN, also a CCF analysis of APM 08279+5255 line and continuum LCs is presented. The results are shown in Fig. \ref{revmapccf}: a peak in the ZDCF appears at $t_{lag}^{ZDCF} = 4095^{+867}_{-369}$ days in the observer frame, simultaneously for the Si {\scriptsize IV} and C {\scriptsize IV}. However, this peak is located in the last significant ZDCF bin, i.e. a bin containing at least the minimal number of points required to compute a statistically significant ZDCF value. The ZDCF point at a longer lag is already not significant. Coupled with the large bin width of the ZDCF peak, this prevents to compute the 68\% probability confidence interval on the peak position over more than one ZDCF bin.

\begin{figure}[htbp]
\begin{center}
\includegraphics[scale=0.45]{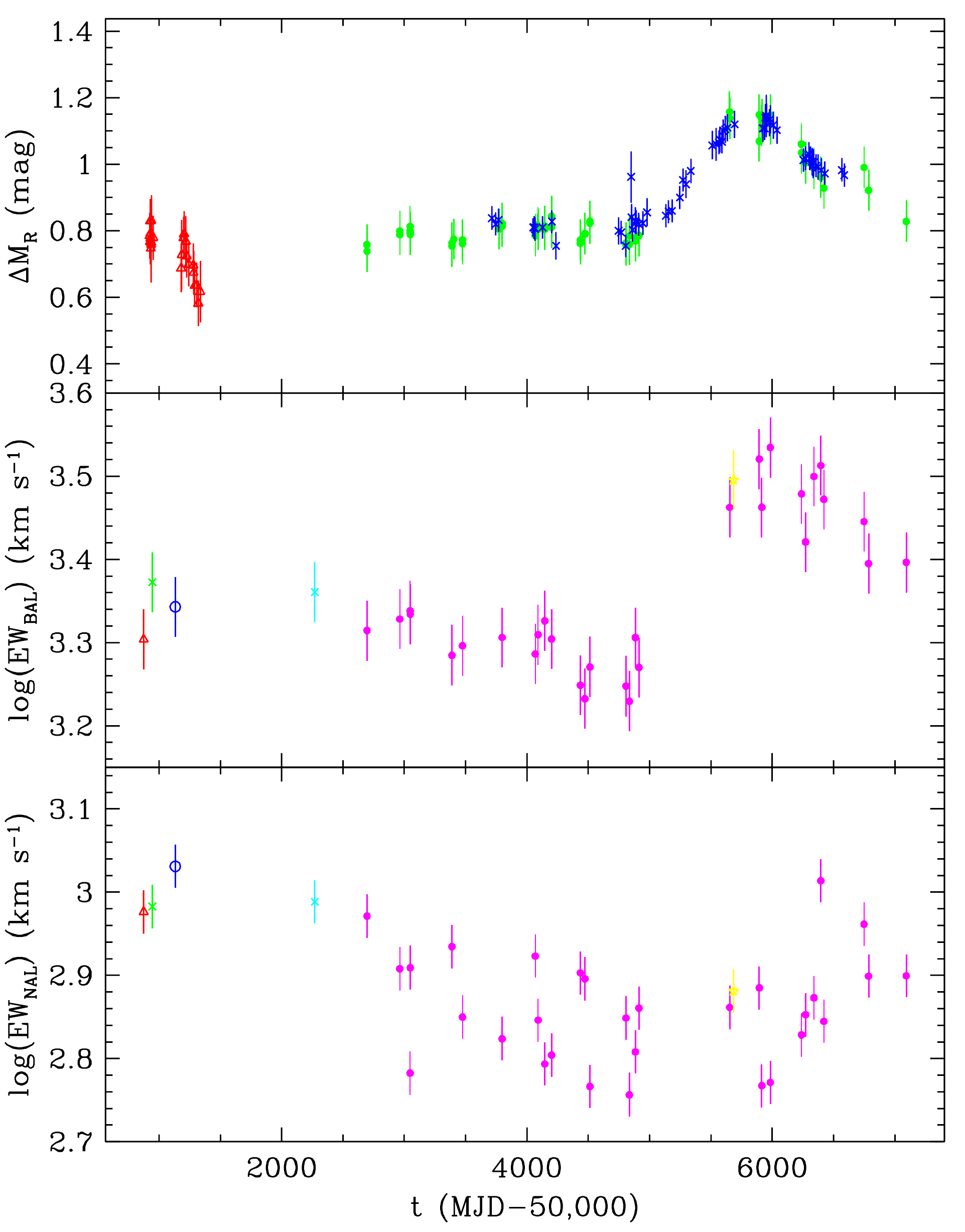}
\end{center}
\caption{{\itshape Upper panel:} updated time series of APM 08279+5255 $R$-band differential magnitude with 1$\sigma$ errors, adding all the photometric data available in literature (same colour code than in the upper panel of Fig. \ref{ligcur}). {\itshape Middle panel:} BAL EWs computed for all the existing APM 08279+5255 spectra with 1$\sigma$ errors. {\itshape Lower panel:} NAL EWs computed for all the existing APM 08279+5255 spectra with 1$\sigma$ errors. In both middle and lower panel, the \citealt{Irw98} ({\itshape red triangle}), the \citealt{Ell99} ({\itshape green cross}), the \citealt{Hin99} ({\itshape blue circle}), the \citealt{Lew02} ({\itshape cyan square}) and the Saturni et al. 2015 ({\itshape yellow star}) spectra are all included along with the Asiago and Loiano spectra ({\itshape magenta dots}).}
\label{baltrend}
\end{figure}

We evaluate $t_{lag}$ from the ICCF estimating the lag uncertainty with the FR/RSS method: we obtain $t_{lag}^{ICCF}\mbox{(Si {\scriptsize IV})} = 4114^{+1414}_{-983}$ days in the observer frame, and $t_{lag}^{ICCF}\mbox{(C {\scriptsize IV})} = 4482^{+1036}_{-1414}$ days. We also quantify the significance of the CCF peaks by performing a Student's $t$-test (\citealt{Bev69}; see also \citealt{She15}), which allows to evaluate the integral probability of the null hypothesis $P(>r, N)$ for a correlation coefficient $r$ computed over $N$ pairs of data points. We compute this probability for both the ICCFs and ZDCFs: for the ICCFs, we obtain $P(>r, N) = 0.03$ for the Si {\scriptsize IV} ($r=0.60$, $N=12$) and 0.05 for the C {\scriptsize IV} ($r=0.55$, $N=12$) respectively; for the ZDCFs, the test gives $P(>r, N) = 0.04$ for the Si {\scriptsize IV} ($r=0.53$, $N=15$) and 0.29 for the C {\scriptsize IV} ($r=0.21$, $N=15$). As expected, the CCFs do not provide a statistically significant result; however, the peaks are all located close to the same position obtained with JAVELIN, indicating that the previously found lags are not an artifact of the adopted model.

\begin{figure}[htbp]
\begin{center}
\includegraphics[scale=0.45]{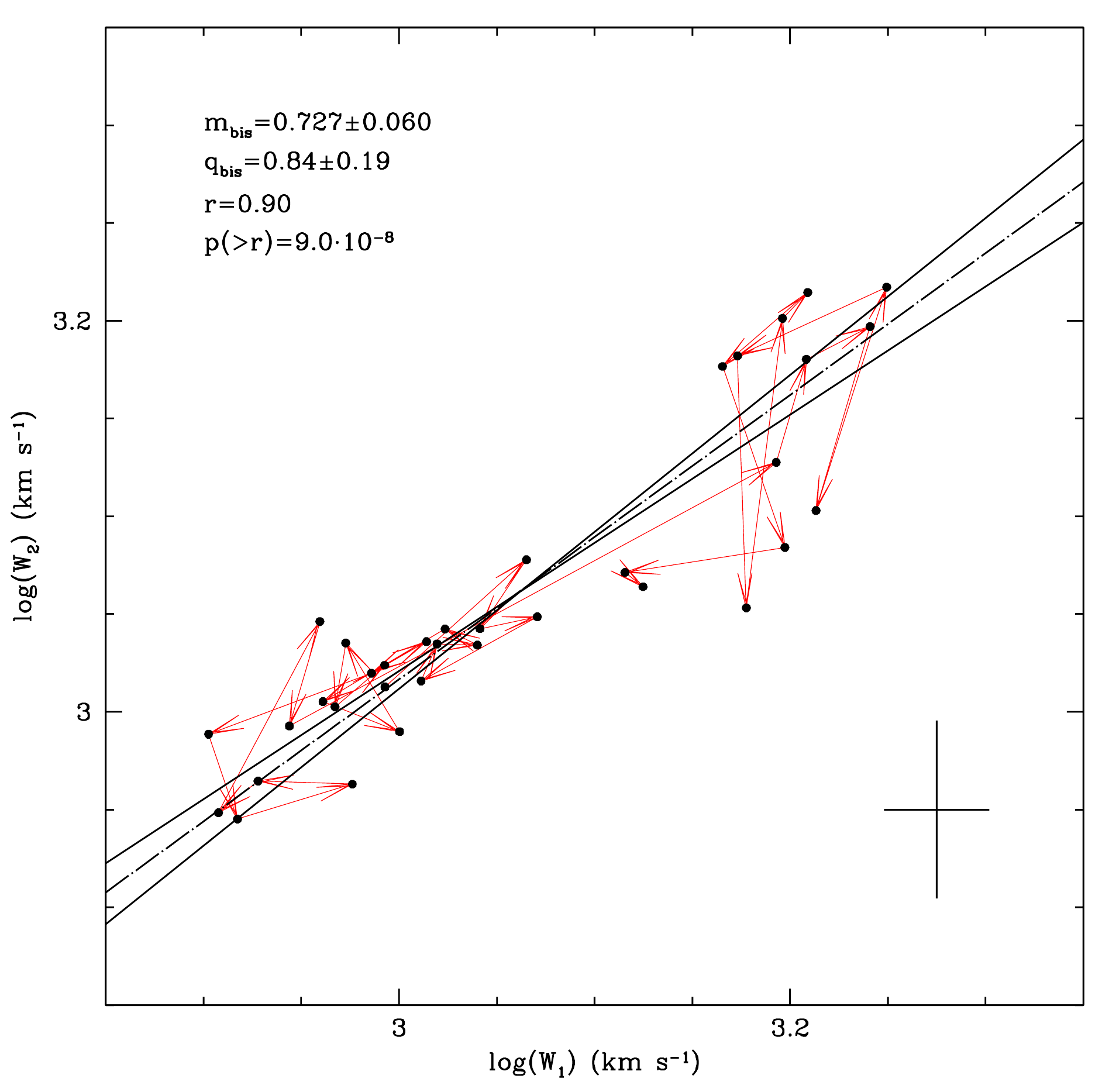}
\end{center}
\caption{EWs of the two C {\tiny IV} BAL troughs of APM 08279+5255, plotted one versus the other ({\itshape black points}). The red arrows connect consecutive epochs of EW pairs. The two correlation fits of BAL$_1$ versus BAL$_2$ and vice versa are shown ({\itshape solid lines}), along with the corresponding bisector fit ({\itshape dot-dashed line}). Slope $m_{bis}$ and intercept $q_{bis}$ of such fit, correlation coefficient $r$ and corresponding null-hypothesis probability $p(>r)$ are reported. The cross in the right-lower corner represents the mean error on the data set in each direction.}
\label{bal12}
\end{figure}

\section{Update of the C {\normalsize IV} absorption variability}\label{avar}

The variability of APM 08279+5255 C {\scriptsize IV} absorption features was presented in Paper I and \citet{Sat14}. In the present paper, we take advantage of the new observations from the Loiano observatory and the photometric data set from the Catalina Sky Survey to extend the absorption variability monitoring in time, and partially fill the gaps in our time series. The data reduction is identical to that performed in Paper I. Fig. \ref{baltrend} shows the update of the C {\scriptsize IV} absorption equivalent width (EW) variability of APM 08279+5255 up to MJD $= 57,093$.

In Paper I, we have identified four absorption systems in the C {\scriptsize IV} region of APM 08279+5255, thanks to the availability of high-resolution spectra. We therefore distinguish two BAL components (respectively labeled BAL$_1$ and BAL$_2$ in Fig. \ref{apmspec}) separated by a NAL system (``blue NAL'' in Fig. \ref{apmspec}) described in \citet{Sri00}. Another NAL system (``red NAL'' in Fig. \ref{apmspec}) is located redwards the C {\scriptsize IV} emission \citep{Sri00,Ell04}. With respect to the previous variability analysis, the ``blue NAL'' has not been included in this update, due to the uncertainty in determining the residual contamination of its EW after the subtraction of the two BAL components. We describe the behaviour of the BAL troughs and the ``red'' NAL in the following subsections.

\subsection{The BAL variability}

\begin{figure*}[htbp]
\begin{center}
\includegraphics[scale=0.7]{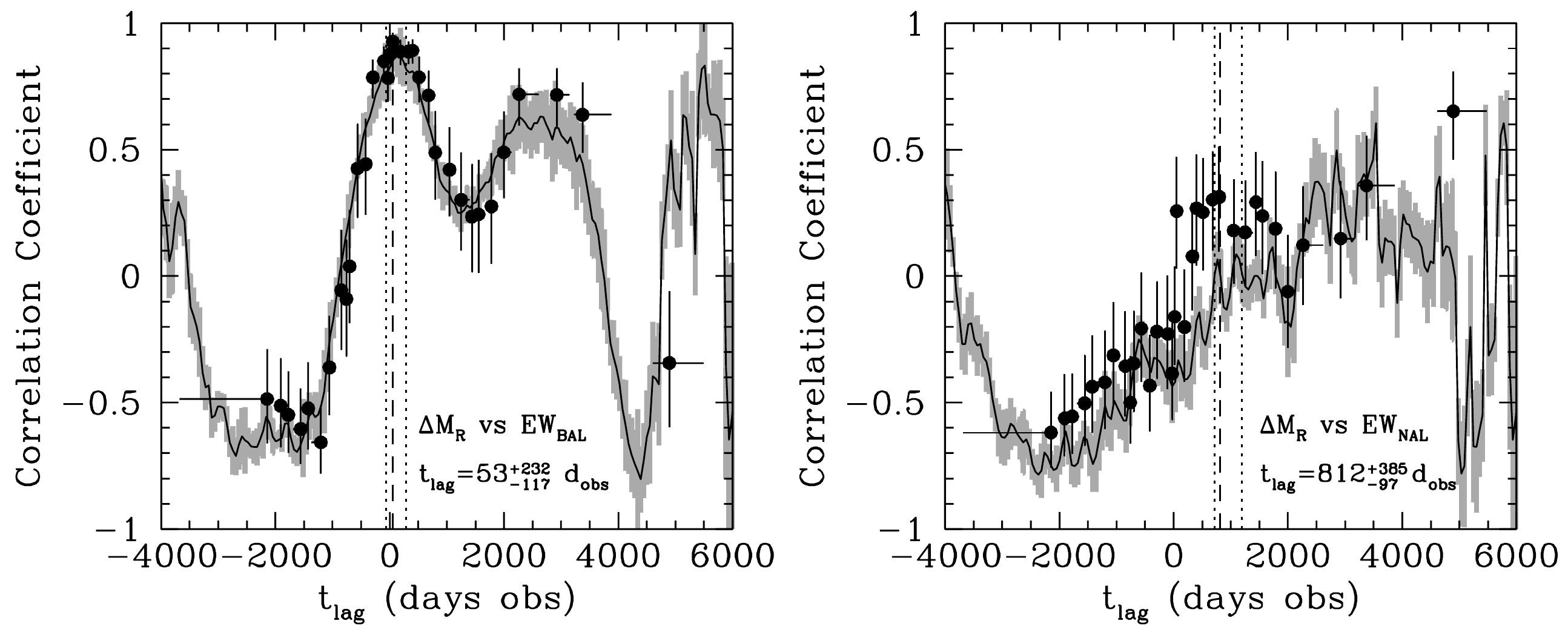}
\end{center}
\caption{{\itshape Left panel:} cross-correlation function of the $R$-band magnitudes with the total C {\tiny IV} BAL EW tracing the continuum. {\itshape Right panel:} cross-correlation function of the $R$-band magnitudes with the total C {\tiny IV} ``red'' NAL EW. In both panels, both the ICCF ({\itshape black solid line}), with 1$\sigma$ errors in the $y$ direction ({\itshape grey bars}), and ZDCF computed with 15 points per bin ({\itshape black points}), with 1$\sigma$ errors in both $x$ and $y$ directions, are shown. The dashed vertical lines mark the delays of the most prominent peaks enclosed by their uncertainties at 68\% probability ({\itshape dotted lines}), which are also numerically reported.}
\label{balcorr}
\end{figure*}

As described in Paper I and \citet{Sat14}, we considered the sum of the two C {\scriptsize IV} BAL components as a single absorption feature for the purpose of the EW variability analysis. This was done since the EWs of such components, computed separately by modeling the BAL profile at low resolution with two Gaussian absorptions, vary in a highly correlated way over the whole monitoring time (see e.g. fig. 5 in Paper I). In order to verify that the new data do not change this result, we show in Fig. \ref{bal12} that this correlation is maintained ($r=0.90, p(>r)=9.0\cdot 10^{-8}$). This result is confirmed by computing the discrete cross-correlation function of the two BAL EWs according to the ZDCF algorithm introduced in Sect. \ref{rmap}, that peaks at $t_{lag} = 171^{+91}_{-513}$ days in the observer frame ($35^{+19}_{-104}$ rest-frame days), which is consistent with no lag between the two components.

\begin{figure}[htbp]
\begin{center}
\includegraphics[scale=0.45]{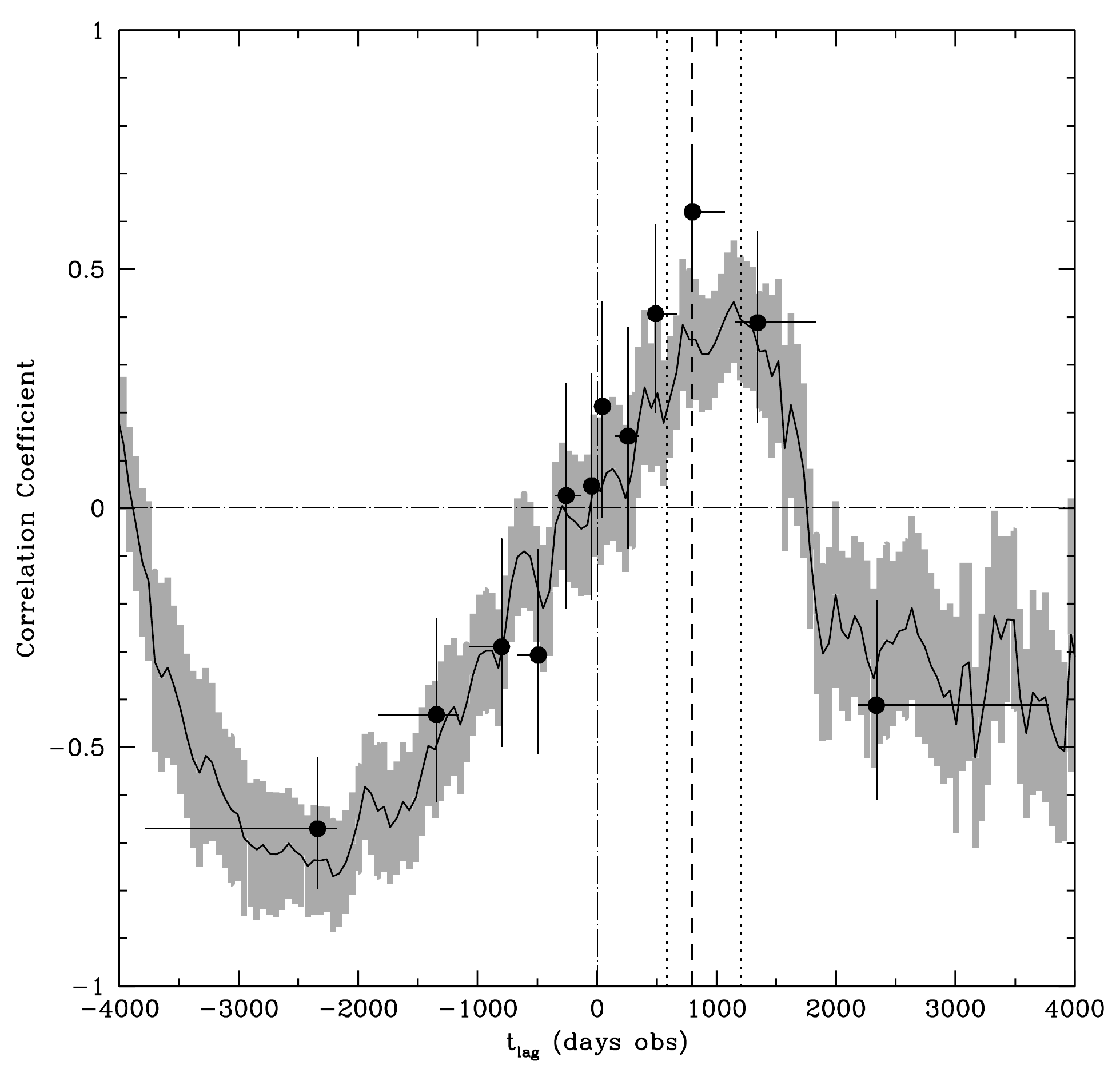}
\end{center}
\caption{Cross-correlation function of the BAL EW with the NAL EW variability. Both the ICCF ({\itshape solid line}), and the ZDCF computed with 20 points per bin ({\itshape black dots}), are shown with their respective 1$\sigma$ errors. The ZDCF lag ({\itshape dashed line}) is indicated, along with its uncertainty at 68\% probability ({\itshape dotted lines}).}
\label{balnalccf}
\end{figure}

This simultaneous variability of different BAL troughs associated to different transitions \citep{Bra14} or the same transition \citep{Gri15} is observed in some other BAL QSOs, indicating common physics driving the BAL variations. We quantify the comparison between BAL$_1$ and BAL$_2$ variability by performing a bisector fit \citep{Akr96} of the relation between the logarithmic BAL EWs $\log{W_1}$ and $\log{W_2}$, obtaining:

\begin{equation}\label{bal1vs2fit}
\log{W_2} = (0.727 \pm 0.060)\log{W_1} + (0.84 \pm 0.19)
\end{equation}

With respect to the simple correlation analyses among $R$-band magnitudes, BAL and NAL EWs presented in Paper I, in this paper we add the cross-correlation analyses. In order to do this, we only apply the ICCF and ZDCF methods, since the reverberation model on which JAVELIN is based cannot be directly applied to absorption variability.

We present the results in the left panel of Fig. \ref{balcorr}: as expected from a visual inspection of Fig. \ref{baltrend}, the correlation peak appears at zero lag ($t_{lag} = 53^{+232}_{-117}$ days in the observer frame, with a significance over 3$\sigma$ level: $N = 15$, $r = 0.94$ and $P(>r, N) = 2 \cdot 10^{-7}$), since the absorbing gas lies along the line of sight and there is no light-travel lag between the continuum and absorption \citep{Bar92,Bar93}. Nonetheless, a physical lag due to recombination time scale could be present, but is however short for high electron densities.

\subsection{The NAL variability}\label{nalvar}

\begin{table*}[htbp]
\centering
\caption{APM 08279+5255 relevant time lags.}
\label{timesc}
\bgroup
\def\arraystretch{1.3}
\begin{tabular}{lccc}
\hline
\hline
\multicolumn{4}{l}{ }\\
Process & Obs-frame lag (days) & Rest-frame lag$^a$ (days) & Method\\
\multicolumn{4}{l}{ }\\
\hline
\multicolumn{4}{l}{ }\\
Si {\scriptsize IV} reverberation (single) & $4095^{+867}_{-369}$ & $834^{+177}_{-75}$ & ZDCF\\
\multicolumn{1}{l}{ } & $4114^{+1414}_{-983}$ & $838^{+288}_{-200}$ & ICCF\\
\multicolumn{1}{l}{ } & $4180^{+1035}_{-323}$ & $851^{+211}_{-66}$ & JAVELIN\\
C {\scriptsize IV} reverberation (single) & $4095^{+867}_{-369}$ & $834^{+177}_{-75}$ & ZDCF\\
\multicolumn{1}{l}{ } & $4482^{+1036}_{-1414}$ & $913^{+211}_{-288}$ & ICCF\\
\multicolumn{1}{l}{ } & $4243^{+337}_{-355}$ & $863^{+69}_{-72}$ & JAVELIN\\
Si {\scriptsize IV} reverberation (joint) & $4286^{+260}_{-417}$ & $873^{+53}_{-85}$ & JAVELIN\\
C {\scriptsize IV} reverberation (joint) & $4425^{+280}_{-560}$ & $901^{+57}_{-114}$ & JAVELIN\\
\multicolumn{4}{l}{ }\\
\hline
\multicolumn{4}{l}{ }\\
BAL$_1$ vs BAL$_2$ variability & $171^{+91}_{-513}$ & $35^{+19}_{-104}$ & ZDCF\\
BAL vs continuum variability & $53^{+232}_{-117}$ & $11^{+47}_{-24}$ & ZDCF\\
NAL vs BAL variability & $799^{+408}_{-214}$ & $163^{+83}_{-44}$ & ZDCF\\
\multicolumn{1}{l}{ } & $1145^{+748}_{-764}$ & $233^{+152}_{-156}$ & ICCF\\
\multicolumn{4}{l}{ }\\
\hline
\multicolumn{4}{l}{{\footnotesize $^a$The rest-frame lag is computed with the systemic redshift $z = 3.911$.}}\\
\end{tabular}
\egroup
\end{table*}

At variance with the BAL behaviour, in Paper I and \citet{Sat14} the ``red NAL'' did not appear to show variations related to the strong continuum variation at MJD $\sim 55,000$. We argued that this could have been caused by a delayed variation in the ionisation state of the NAL absorber, not yet seen in the LC, with respect to the continuum changes. We estimated a lower limit to this physical recombination time lag of $\sim 200$ rest-frame days, roughly corresponding to the time span between MJD $\sim 55,000$ and $\sim 56,000$ in our data (see Fig. \ref{baltrend}). With this limit, we placed an upper limit of $n_e \lesssim 2\cdot 10^4$ cm$^{-3}$ on the electron density of the absorber.

With the update of the NAL EW time series presented in Fig. \ref{baltrend}, a delayed rise seems to appear with respect to the continuum variation of $\sim 0.4$ mag after MJD $\sim 55,000$, with the NAL EW rising by $\sim 0.2$ dex between MJD $\sim 56,000$ and 57,000. At a glance, this would correspond to a rest-frame time lag of roughly 200 days. A measurement of this delay can provide a more robust estimate of the NAL electron density instead of the upper limit of Paper I.

We therefore compute the CCFs between the NAL EW and continuum variability. Both the ICCF and ZDCF are shown in the right panel of Fig. \ref{balcorr}: at variance with the BAL case, here the correlation peak appears at different lags for ICCF ($3541^{+1090}_{-2424}$ days in the observer frame) and ZDCF ($812^{+385}_{-97}$ days), and both peaks are marginally significant at 2$\sigma$ level ($N = 19$, $r = 0.60$, $P(>r, N) = 0.01$ for the ICCF; $N = 15$, $r = 0.65$, $P(>r, N) = 0.01$ for the ZDCF). Also a visual inspection of Fig. \ref{baltrend} shows that the identification of the main CCF peaks is not straightforward.

However, in Paper I we suggested that the absorption variability occurs in response to the variations in the {\itshape true} C {\scriptsize IV} ionising continuum at $\sim 200$ \AA, which is not observed. The BAL variability is assumed to trace this ionising continuum with no or very small recombination lag due to a high electron density of the absorbing wind. Therefore, we assume the BAL EW time series as a proxy of the ionising continuum variability, and compute ICCF and ZDCF with the NAL EWs. The result is shown in Fig. \ref{balnalccf}: a single peak arises at $t_{lag} = 1145^{+748}_{-764}$ days in the observer frame for the ICCF and at $t_{lag} = 799^{+408}_{-214}$ days for the ZDCF, with a significance at more than 2$\sigma$ level in both cases ($N = 26$, $r = 0.43$, $P(>r, N) = 0.02$ for the ICCF; $N = 20$, $r = 0.62$, $P(>r, N) = 3 \cdot 10^{-3}$ for the ZDCF). This favours the correlation between the continuum flux decrease beginning at MJD $\sim 55,000$ and the NAL EW rise after MJD $\sim 56,000$. Moreover, it suggests at the same time that the BAL variability is in fact a better proxy of the ionising continuum variations than the $R$-band continuum.

\section{Discussion}\label{disc}

Thanks to our long-time spectro-photometric monitoring of the high-$z$ bright quasar APM 08279+5255, we have obtained several time lags linking the variability of emission and absorption lines to that of the driving continuum emission. In this way, we are able to probe (i) the region of the ionised gas clouds producing the emission lines, and (ii) the regions where the high-velocity outflows responsible of broad absorption in quasar spectra originate. In Tab. \ref{timesc}, we summarise all these time lags together, giving them both in the observer frame and in the rest frame for $z = 3.911$.

In this section, we discuss what can be inferred about the relevant physical parameters of these regions. From the Si {\scriptsize IV} and C {\scriptsize IV} emission-line lags, we can estimate the black hole mass of APM 08279+5255 with some assumptions on the BLR shape and inclination (which are summarised into the form factor $f$). From the C {\scriptsize IV} NAL recombination lag with respect to the ionising continuum at $\sim 200$ \AA, we can obtain information about the electron density and the distance of the absorbing gas from the central engine.

\subsection{Si {\small IV} and C {\small IV} BLR stratification}

Since the time lags obtained for Si {\scriptsize IV} and for C {\scriptsize IV} are equal within errors, this means that the respective BLRs must be approximately located at the same distance from the central black hole. How does this compare with the BLR size ratio found in low-luminosity AGNs? In order to explore this point, we collected from the literature all the objects having RM time lags of both Si {\scriptsize IV} and C {\scriptsize IV}. These objects are the Seyfert galaxies NGC 3783, NGC 5548 and NGC 7469 \citep{Pet04}.

\begin{figure}[htbp]
\begin{center}
\includegraphics[scale=0.45]{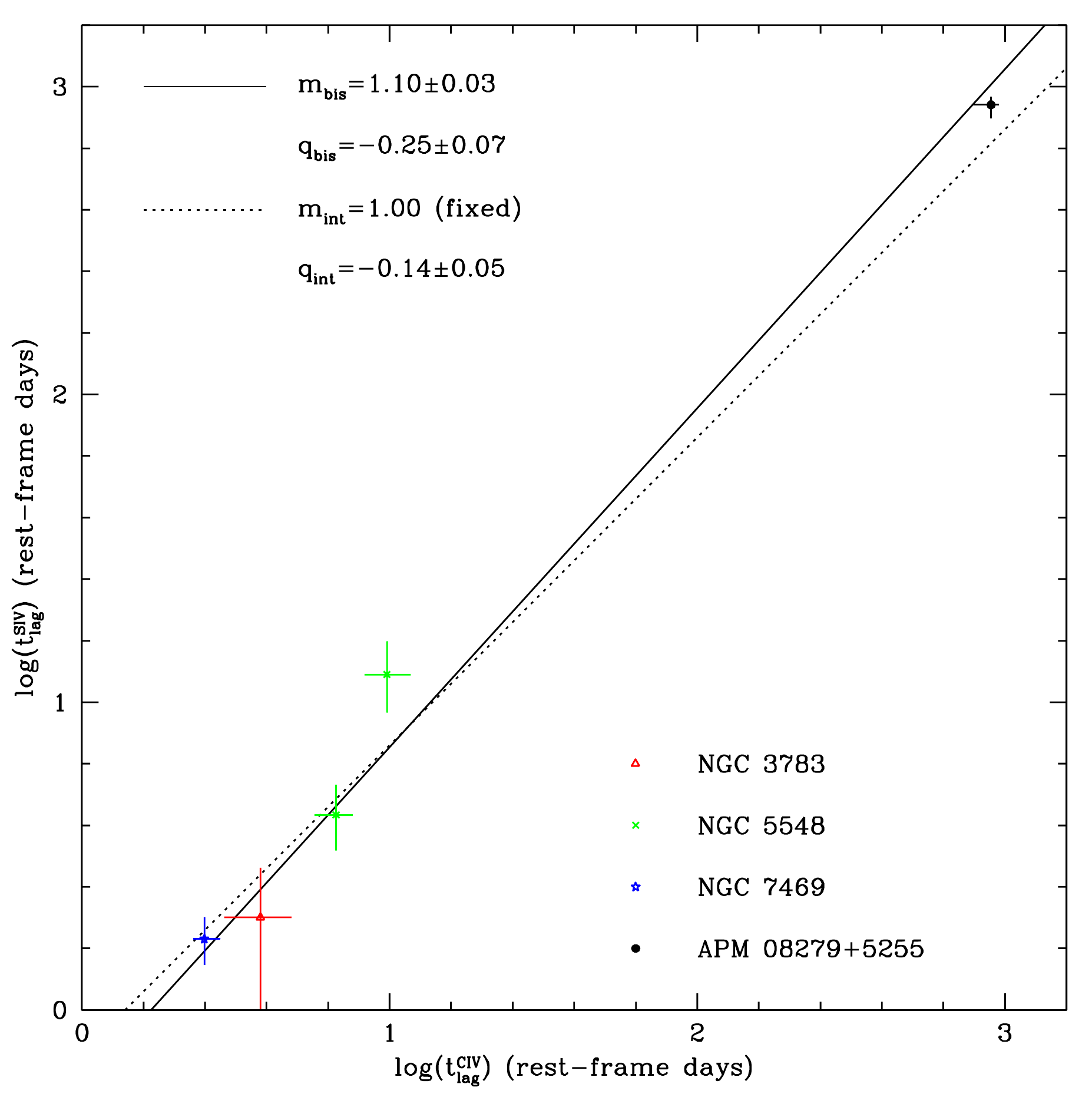}
\end{center}
\caption{Rest-frame Si {\tiny IV} time lags as a function of C {\tiny IV} lags. All the available data from the literature are shown, together with our measurements from the RM of APM 08279+5255 (legend is on the plot). For the whole data set, both a free linear fit in the logarithmic space ({\itshape solid line}), and a linear fit with logarithmic slope fixed to unity ({\itshape dotted line}), have been performed. Fit coefficients are indicated in the plot with their $1\sigma$ errors.}
\label{blr}
\end{figure}

In Fig. \ref{blr} we show such collection of rest-frame time lags from Si {\scriptsize IV} as a function of those from C {\scriptsize IV}, and perform a logarithmic bisector fit to the data in order to quantify the average size difference between Si {\scriptsize IV} and C {\scriptsize IV} BLRs from Seyfert galaxies to quasars, spanning three orders of magnitude of BLR size. Allowing both slope and intercept to vary, the fit indicates a possible increase of the time lag ratio, and hence of the BLR size ratio, with a power-law behaviour with logarithmic slope $1.10 \pm 0.03$. For comparison, a fit with a unitary slope is also shown in Fig. \ref{blr}, corresponding to an average ratio $\langle R_{BLR}^{SiIV}/R_{BLR}^{CIV} \rangle = \langle t_{lag}^{SiIV}/t_{lag}^{CIV} \rangle = 0.72^{+0.10}_{-0.08}$. This result is dominated by the Seyfert galaxies, whose average ratio is $\langle R_{BLR}^{SiIV}/R_{BLR}^{CIV} \rangle = 0.69^{+0.11}_{-0.10}$.

\subsection{The size-luminosity relation for the C {\small IV} BLR}

\begin{figure}[htbp]
\begin{center}
\includegraphics[scale=0.45]{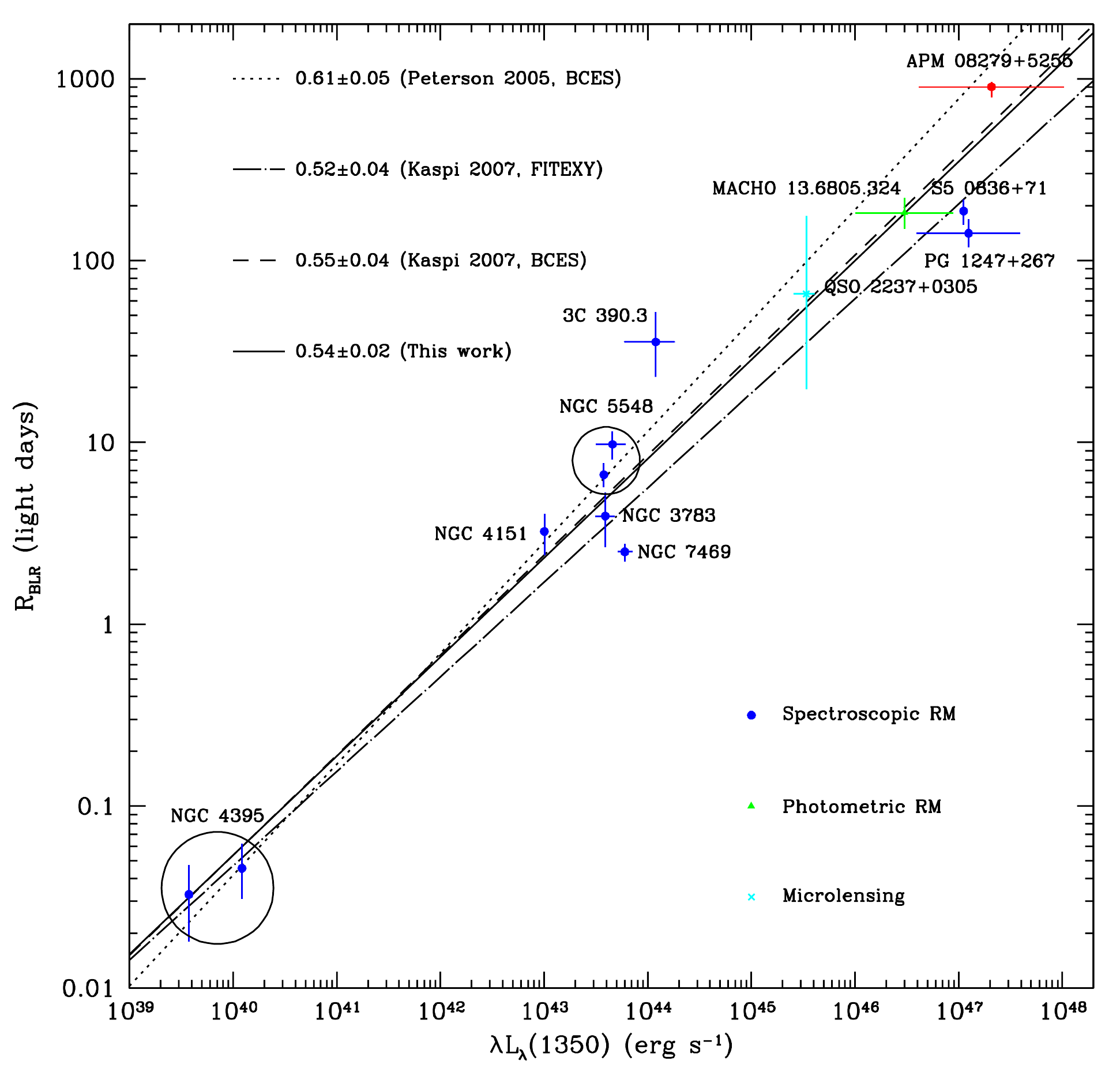}
\end{center}
\caption{Size-luminosity relation for C {\tiny IV} BLR. Together with APM 08279+5255, all the objects with measurements of C {\tiny IV} BLR size and monochromatic luminosity at 1350 \AA~available in the literature have been included in this plot: NGC 4395, NGC 7469, NGC 3783, NGC 5548 and 3C 390.3 from \citet{Pet04,Pet05}, NGC 4151 from \citet{Met06}, QSO 2237+0305 from \citet{Slu11}, MACHO 13.6805.324 from \citet{Che12}, S5 0836+71 from \citet{Kas07} and PG 1247+267 from \citet{Per14,Tre14}. Colour and symbol codes are explained in the plot. For comparison with our fit, also the fits performed by \citet{Pet05,Pet06} and \citet{Kas07} are reported (see legend in the plot).}
\label{rlciv}
\end{figure}

Our result for APM 08279+5255 allows us to extend the size-luminosity relation for the C {\scriptsize IV} emitting region up to $\sim 10^{48}$ erg s$^{-1}$ in monochromatic luminosity at 1350 \AA~$\lambda L_\lambda (1350)$. In Fig. \ref{rlciv} we collect al the available data for the size of the C {\scriptsize IV} emitting region in AGNs as a function of their $\lambda L_\lambda (1350)$, where we add our point for APM 08279+5255 to spectroscopic RM measurements \citep{Pet04,Pet05,Met06,Kas07,Per14,Tre14}. Also, we add the point for the quasar MACHO 13.6805.324 obtained with photometric RM by \citet{Che12}, and the point for the gravitationally lensed quasar QSO 2237+0305 obtained through microlensing by \citet{Slu11}.

We fit a power-law relation $R_{BRL} \propto L_{1350}^\beta$ to this data set with a method that takes into account the uncertainties in both variables and the intrinsic scatter of points. We run a high number of bisector fits \citep{Akr96} on data sets in which each original point is replaced by casting a pair of substituting values within the associated error box; these values are weighted by an asymmetric distribution function, represented by two demi-Gaussian distributions with different standard deviations, corresponding to the asymmetric uncertainties. The best-fit relation obtained with $10^4$ runs is:

\begin{equation}\label{rlciveq}
\log{R_{BLR}} = (0.9 \pm 0.7) + (0.54 \pm 0.02) \log{\left[
\frac{\lambda L_\lambda (1350)}{10^{44}\mbox{ erg s}^{-1}}
\right]}
\end{equation}

\begin{figure*}[htbp]
\begin{center}
\includegraphics[scale=0.7]{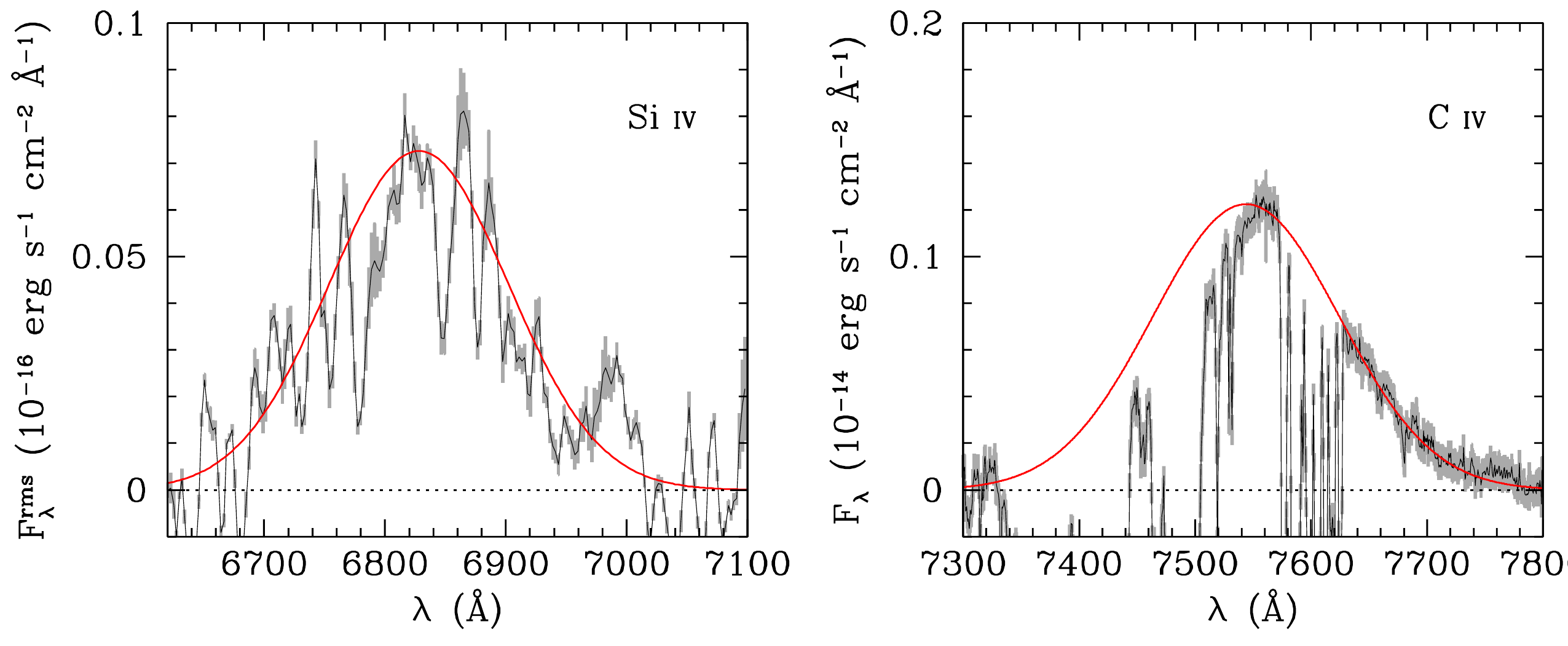}
\end{center}
\caption{{\itshape Left panel:} continuum-subtracted rms spectrum of APM 08279+5255 Si {\tiny IV} emission line from our RM monitoring. {\itshape Right panel:} continuum-subtracted HST spectrum of the C {\tiny IV} emission line. In both panels, errors on the flux level ({\itshape grey bars}) and best-fit Gaussians ({\itshape red curves}) are shown. For Si {\tiny IV}, the major deviations from the best-fit curve are the residuals of intrinsic narrow absorptions superimposed to the emission line, as can be seen from the analysis by \citet{Sri00} and \citet{Ell04}.}
\label{siglin}
\end{figure*}

With respect to previous relations \citep{Pet05,Kas07}, having added objects with uncertain luminosities to the fit gives a large error in the intercept. In the two cases of the quasars PG 1247+267 and APM 08279+5255, this is due to the unknown lens magnification factor $\mu$. In fact, as described in \citet{Tre14}, the first object has relatively small emission-line widths with respect to its luminosity, an Eddington ration of $\sim 10$, and a significant deviation from the $\alpha_{OX}$ -- $L_{2500}$ relation \citep{She14}: an invisible gravitational lens with $\mu \approx 10$ could explain at once all the anomalies described in \citet{Tre14}. For APM 08279+5255, instead, the fact that the lensing galaxy is not visible \citep{Lew02} leads to very model-dependent estimates of the magnification factor. Currently, $\mu$ ranges between $\sim 100$ \citep{Ega00} and $\sim 4$ \citep{Rie09} for APM 08279+5255; therefore, in order to provide a rough indication of the global uncertainty due to both the statistical errors and the ignorance of the lens magnification, we adopt as uncertainty in the quasar luminosity the whole range of plausible $\lambda L_\lambda (1350)$ between $\mu = 4$ and 100 for APM 08279+5255, and the range between $\mu = 1$ (i.e. no magnification) and 10 for PG 1247+267.

\subsection{APM 08279+5255 black hole mass estimates}\label{bhmass}

In order to give a direct estimate of the black hole mass of APM 08279+5255, we follow the approach of \citet{Tre14}. Therefore, we first identify the velocity $\Delta v$ appearing in Eq. \ref{revmapeq} with the rms velocity dispersion along the line of sight $\sigma_l$. We then compute the uncertainty on $\sigma_l$ for Si {\scriptsize IV} by applying the bootstrap method described in \citet{Pet04}, obtaining $\Delta v_{SiIV} = 3245^{+101}_{-107}$ km s$^{-1}$; the use of $\sigma_l$ both avoids underestimates of $\Delta v$ caused by narrow emission-line components and the effect of non-virial outflows varying on time scales longer than RM lags (e.g., \citealt{Den12}).

Unfortunately, we cannot apply the same procedure to C {\scriptsize IV}, since its emission-line profile is heavily affected by absorption features that prevent to compute a reliable rms spectrum in that region. Therefore, instead of using $\sigma_l$, for C {\scriptsize IV} we adopt the standard deviation of a Gaussian profile $\sigma_g = 3200 \pm 30$ km s$^{-1}$, as measured through the C {\scriptsize IV} emission profile fit performed on the high-resolution dereddened HST/STIS spectrum of APM 08279+5255. Since the unabsorbed spectral intervals in this spectrum appear to be free of narrow components and outflows in emission, we can consider this as a good equivalent estimate of the proper C {\scriptsize IV} rms $\sigma_l$. In Fig. \ref{siglin}, we show both the fit of C {\scriptsize IV} on the HST spectrum and a Gaussian fit of the rms Si {\scriptsize IV}, performed by only allowing the rms line amplitude to vary, setting the peak wavelength to $\lambda$1400 \AA~at $z = 3.87$ \citep{Irw98,Dow99} and the line width to the Si {\scriptsize IV} $\sigma_l$.

\begin{figure}[htbp]
\begin{center}
\includegraphics[scale=0.45]{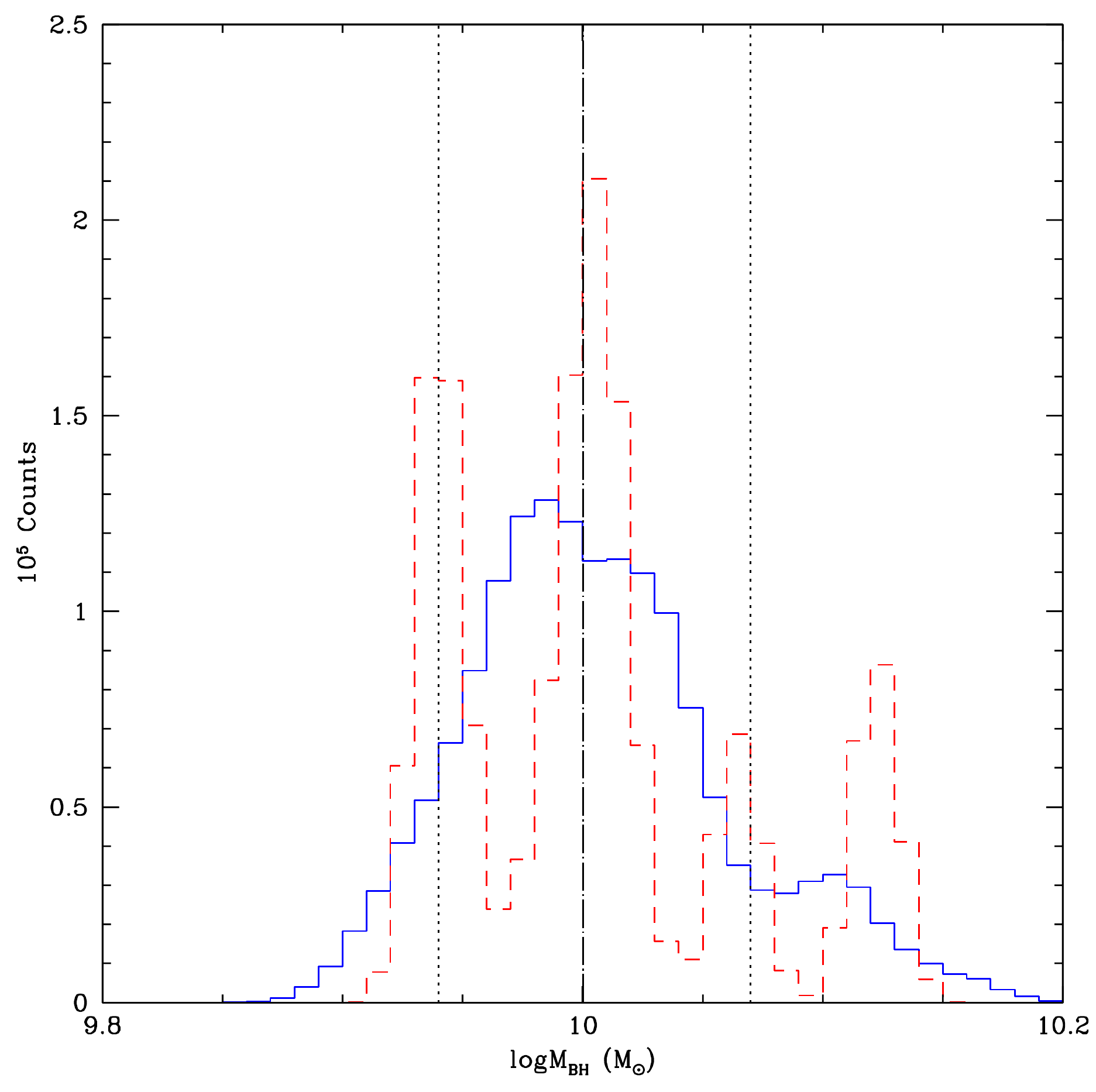}
\end{center}
\caption{Posterior distribution of APM 08279+5255 black hole mass from Si {\tiny IV} ({\itshape blue solid histogram}) and C {\tiny IV} ({\itshape red dashed histogram}) RM. The best estimate of the black hole mass ({\itshape dot-dashed line}) is indicated together with its confidence intervals at 68\% probability ({\itshape dotted lines}).}
\label{mbhhpd}
\end{figure}

Finally, we multiply the posterior distributions of $t_{lag}$ for Si {\scriptsize IV} and C {\scriptsize IV} for the respective statistical distributions of $\Delta v^2$ to obtain two posterior distributions of APM 08279+5255 virial products $t_{lag} \Delta v^2$, to be inserted in Eq. \ref{revmapeq} in order to obtain the black hole mass $M_{BH}$. We adopt as form factor the value $f = 5.5$, obtained by \citet{Onk04} through the calibration of the H$\beta$ RM masses on the relation between black hole masses and stellar velocity dispersions in galaxy bulges \citep{Fer00,Mer01b,Gul09}, which is appropriate for the definition of $\Delta v = \sigma_l$ (not FWHM); although more recent estimates provide different values for $f$ (e.g., \citealt{Pan14} and refs. therein), we choose this commonly used value for a direct comparison with the literature nonetheless, as also done in \citet{Tre14}. Fig. \ref{mbhhpd} shows such distributions, that give an identical result of $M_{BH} = \left(1.00^{+0.17}_{-0.13}\right) \cdot 10^{10}$ $M_\odot$ as best estimate of the virial mass.

\subsection{Estimate of the lens magnification}\label{maglens}

APM 08279+5255 is a well-known gravitational lens, first confirmed case with an odd number of images of the lensed source and no observed trace of the lensing galaxy \citep{Lew02}. Therefore, the estimate of the lens magnification $\mu$ only relies on modeling the lensed quasar image from high-resolution optical/infrared imaging, and the values of $\mu$ are significantly discrepant when estimated from different lens models \citep{Ega00,Rie09}. This primarily affects the single-epoch black hole mass estimates from mass-luminosity virial relationships, that can give differences in the value of $M_{BH}$ of up to one order of magnitude when adopting such magnifications.

Our direct RM measurement of the C {\scriptsize IV} $R_{BLR}$ for APM 08279+5255 allows us to invert the relation between the BLR size and UV luminosity, thus giving a model-independent estimate of $\mu$ that can discriminate between the competing models. We thus solve eq. 3 of \citet{Kas07} for $\lambda L_\lambda (1350)$, which is not affected by our ignorance on APM 08279+5255 true luminosity, adopting our fiducial JAVELIN time lag for C {\scriptsize IV} of $901$ days in the rest frame:

\begin{equation}\label{sembh}
\lambda L_\lambda^{true} (1350) = 10^{42.56} R_{BLR}^{1.92} = \left(
1.7^{+6.6}_{-1.2}
\right) \cdot 10^{48} \mbox{ erg s}^{-1}
\end{equation}

With a measured luminosity $\lambda L^{meas}_\lambda (1350) = 4.13 \cdot 10^{48}$ erg s$^{-1}$ from our de-reddened spectra, the lens magnification can be estimated by:

\begin{equation}
\mu = \frac{\lambda L_\lambda^{meas}(1350)}{\lambda L_\lambda^{true}(1350)} = 2.4^{+5.8}_{-1.9}
\end{equation}

Although affected by large errors, due to the uncertainty on the BLR-luminosity relation, such a low magnification favours the lens model by \citet{Rie09} with respect to that by \citet{Ega00}.

\subsection{Black hole mass versus host-galaxy mass: an overmassive black hole for APM 08279+5255}\label{over}

Having measured the black hole mass of APM 08279+5255, we now address the question of how it compares with the mass of the host galaxy. From the known relation between black hole mass and host-galaxy bulge mass for nearby galaxies \citep{Mag98,Mer01a,Mar03,Har04}, the expected bulge mass $M_{bulge}$ for APM 08279+5255 would be $\sim 400$ times $M_{BH}$; from CO emission-line structure, \citet{Rie09} derived $M_{bulge} = 3.0 \cdot 10^{11}/\mu$ $M_\odot$, that corresponds to $7.5 \cdot 10^{10}$ $M_\odot$ for $\mu = 4$. Using our RM measurement of $M_{BH}$, this means that $M_{bulge} = 7.5 M_{BH}$ for APM 08279+5255, slightly higher than \citet{Rie09} estimate but still more than 50 times lower than the value extrapolated from the local $M_{bulge}-M_{BH}$ ratio.

This revised result confirms the conclusions by \citet{Rie09}, i.e. a black-hole mass assembly already ended at very early cosmic times into a galaxy that has still to form the largest part of its stars. A handful of other objects presenting a similar $M_{BH}/M_{bulge} \sim 0.1$ are known so far \citep{Wal04,Rie08a,Rie08b}, APM 08279+5255 being the first one with a direct estimate of its $M_{BH}$. High $M_{BH}/M_{bulge}$ ratios may be common at high redshift, with a non-evolving black hole-to-{\itshape total} stellar mass ratio ($M_{BH}/M_*$) as stars later settle into galaxy bulges and a $M_{BH}/M_{bulge}$ ratio at $z \sim 0$ (e.g., \citealt{Jah09}).


The value of $M_*$ for APM 08729+5255 is currently unknown; still, we note that, in the $M_{BH} - M_{*}$ diagram (see fig. 2 of \citealt{Tra15}), APM 08279+5255 hosts one of the most massive black holes measured so far. Its $M_{BH}$ is in fact slightly higher than CID-947, a $z \sim 3.3$ quasar interpreted as a possible prototype of a class of objects with still largely incomplete star formation in the host galaxy \citep{Tra15}.

\subsection{Physics and location of the C {\small IV} absorbing gas}\label{absphys}

We argued in Paper I that the most plausible mechanism for absorption variability in APM 08279+5255 is a change in the photo-ionisation state of the gas driven by the variability of the ionising continuum. Within this framework, the increase in the absorption EWs after the flux drop at MJD $\sim 55,000$ is due to an equivalent increase in the density of C {\scriptsize IV} ions, through the recombination process C$^{4+} \rightarrow$ C$^{3+}$. The time scales obtained through the CCF between the absorption and continuum variability measure the delay between a variation in the quasar radiation field and the corresponding variation in the density of the absorbing ions, through which we can infer the density of the absorbing gas, and thus its distance from the central engine, or at least a lower limit.
 
Since the change in the BAL EW is simultaneous with the $R$-band strong continuum variation, the time lag obtained for the main C {\scriptsize IV} broad absorption features is compatible with zero. Thus, we cannot establish a significant lower limit to the electron density of the BAL wind. 

Conversely, the observed lag in the NAL EW variability can give an estimate of the density of the absorbing clouds $n_e$. Assuming $\alpha_{rec} = 2.8 \cdot 10^{-12}$ cm$^{3}$ s$^{-1}$ \citep{Arn85} as the recombination coefficient of C$^{4+}$ atoms, we compute the value of $n_e$ as follows:

\begin{equation}\label{nenal}
n_e = \frac{1}{t_{lag}\alpha_{rec}} = \left(
2.5^{+1.0}_{-0.8}
\right) \cdot 10^4 \mbox{ cm}^{-3},
\end{equation}
assuming the ZDCF lag computed with respect to the BAL variability (see Sect. \ref{nalvar}) as an estimate of the NAL recombination time lag.

A rough estimate of the distance of the NAL gas from the continuum source may be obtained through the knowledge of the ionisation parameter $U$ of APM 08279+5255, which is linked to the distance from the ionising source and to its luminosity by (e.g., \citealt{Pet97}):

\begin{equation}\label{ionrate}
U = \frac{1}{4\pi r^2 c n_e} \int_{\nu_{ion}}^{+\infty} \frac{L_\nu}{h\nu} d\nu
\end{equation}

We can derive $r$ as:

\begin{equation}\label{req}
r_{NAL} = \sqrt{\frac{1}{4\pi U c n_e}\int_{\nu_{ion}}^{+\infty} \frac{L_\nu}{h\nu} d\nu},
\end{equation}
where we adopt $U \sim 10^{-2}$ from \citet{Sri00}. In order to estimate $L_\nu$ over the range of (unobserved) ionising energies, we follow the procedure of \citet{Gri15}, calibrating the synthetic quasar spectral energy distribution (SED) of \citet{Dun10} to APM 08279+5255 bolometric luminosity $L_{bol} \sim 4.5\cdot 10^{48}$ erg s$^{-1}$. Such luminosity is inferred from the observed flux density at 3000 \AA~of the optical/IR spectrum of APM 08279+5255 taken at the Telescopio Nazionale Galileo (La Palma, Canarian Islands; Saturni et al., in prep.), with a bolometric correction factor of 5 \citep{Ric06} and the lens magnification $\mu = 4$ of \citet{Rie09}; with this normalisation, we compute $\int_{\nu_{ion}}^{+\infty}\left(L_\nu/h\nu\right) d\nu = 8.2 \cdot 10^{58}$ s$^{-1}$ for energies of up to 50 keV. Substituting these quantities in Eq. \ref{req}, we finally obtain $r_{NAL} \approx 9.6$ kpc, compatible with a location of the NAL absorbers in the body of APM 08279+5255 host galaxy.

\section{Conclusions}\label{conc}

We can summarise our results as follows.

\begin{enumerate}
\item The long monitoring of APM 08279+5255 obtained by combining our observational campaign with literature data, with a total sample of 138 photometric $R$-band points and 30 spectra (plus 5 from the literature, used for the study on the C {\scriptsize IV} absorption variability) spanning $\sim 16$ years in the observer frame ($\sim 3.5$ years in the rest frame), allowed us to perform the RM of this quasar. The resulting rest-frame time lag is of $\sim 900$ days in the rest frame for both Si {\scriptsize IV} and C {\scriptsize IV}.

\item We find $t_{lag}^{SiIV}/t_{lag}^{CIV} \sim 1$ for APM 08279+5255. This ratio is only marginally consistent with the average value found for Seyfert galaxies ($\langle t_{lag}^{SiIV}/t_{lag}^{CIV} \langle = 0.7 \pm 0.1$), thus possibly indicating a slight increase of $t_{lag}^{SiIV}/t_{lag}^{CIV}$ from Seyferts to quasars.

\item We update the distance-luminosity relation for quasars obtained through C {\scriptsize IV} lags \citep{Pet05,Kas07}, and compute a black hole mass of $\sim 10^{10}$ $M_\odot$ for APM 08279+5255.

\item With our direct time lag measurement, we can invert the distance-luminosity relation for C {\scriptsize IV} of \citet{Kas07} in order to obtain an estimate of APM 08279+5255 lens magnification, so far uncertain between 100 \citep{Ega00} and 4 \citep{Rie09} depending on the model of the lensing galaxy shape, size and positioning. This provides an estimate of $\mu = 2.4^{+5.8}_{-1.9}$, consistent at $1\sigma$ level with the model of \citet{Rie09} and disfavouring the model of \citet{Ega00} instead.

\item We revise the $M_{BH}/M_{bulge}$ ratio presented by \citet{Rie09}, confirming that APM 08279+5255 has an oversized black hole with respect to its host-galaxy bulge \citep{Wal04,Rie08a,Rie08b}. A future measurement of its host-galaxy stellar mass would be extremely important in order to establish whether APM 08279+5255 host galaxy has an under-developed stellar component, in analogy with the case of CID-947 \citep{Tra15}.

\item We update the study on the C {\scriptsize IV} absorption variability already presented in Paper I and \citet{Sat14}, and further strengthen the hypothesis of BAL variations driven by changes in the C {\scriptsize IV} ionising continuum at $\sim 200$ \AA.

\item At variance with the BAL components, the narrow absorption system redwards the C {\scriptsize IV} $\lambda 1549$ emission exhibits a variation delayed by $\sim 160$ days in the rest frame with respect to the ionising continuum. Under the assumption that this lag is due to C$^{+4} \rightarrow$ C$^{+3}$ recombination, we estimate a distance of the NAL gas from the central engine consistent with galactic sizes.

\end{enumerate}

\begin{acknowledgements}
We thank our anonymous referee for their helpful comments. We are also grateful to Ying Zu (McWilliams Center for Cosmology, Carnegie Mellon University) for the useful discussion about the use of JAVELIN. We acknowledge funding from PRIN/MIUR-2010 award 2010NHBSBE. M.D. acknowledges PRIN INAF 2011 funding. This research is based on observations collected at the Copernico telescope (Asiago, Italy) of the INAF-Osservatorio Astronomico di Padova, and at the Cassini Telescope (Loiano, Italy) of the INAF-Osservatorio Astronomico di Bologna.
\end{acknowledgements}

\bibliographystyle{aa}
\bibliography{biblio}

\newpage

\begin{table*}[htbp]
\centering
\caption{\label{tabflux} APM 08279+5255 continuum and emission-line fluxes.}
\begin{tabular*}{\textwidth}{c @{\extracolsep{\fill}} cccccc}
\hline
\hline
\multicolumn{6}{l}{ }\\
MJD$-50,000$ (days) & Telescope & Obs Type & $F_Q/F_S$ & $F_{\mbox{\scriptsize SiIV}}$ ($10^{-13}$ erg s$^{-1}$ cm$^{-2}$) & $F_{\mbox{\scriptsize CIV}}$ ($10^{-13}$ erg s$^{-1}$ cm$^{-2}$)\\
\multicolumn{6}{l}{ }\\
\hline
\multicolumn{6}{l}{ }\\
925.0 & V$^1$ & P$^a$ & $0.485 \pm 0.031$ & & \\
928.0 & V & P & $0.466 \pm 0.029$ & & \\
930.0 & V & P & $0.493 \pm 0.031$ & & \\
932.0 & V & P & $0.482 \pm 0.030$ & & \\
933.0 & V & P & $0.488 \pm 0.032$ & & \\
934.0 & V & P & $0.502 \pm 0.033$ & & \\
936.0 & V & P & $0.494 \pm 0.035$ & & \\
937.0 & V & P & $0.497 \pm 0.052$ & & \\
938.0 & V & P & $0.466 \pm 0.033$ & & \\
953.0 & V & P & $0.488 \pm 0.030$ & & \\
1181.0 & V & P & $0.531 \pm 0.035$ & & \\
1186.0 & V & P & $0.512 \pm 0.050$ & & \\
1202.0 & V & P & $0.488 \pm 0.029$ & & \\
1205.0 & V & P & $0.483 \pm 0.031$ & & \\
1218.0 & V & P & $0.493 \pm 0.035$ & & \\
1227.0 & V & P & $0.513 \pm 0.031$ & & \\
1243.0 & V & P & $0.525 \pm 0.032$ & & \\
1281.0 & V & P & $0.527 \pm 0.032$ & & \\
1282.0 & V & P & $0.537 \pm 0.032$ & & \\
1292.0 & V & P & $0.557 \pm 0.033$ & & \\
1307.0 & V & P & $0.557 \pm 0.034$ & & \\
1320.0 & V & P & $0.585 \pm 0.037$ & & \\
1337.0 & V & P & $0.566 \pm 0.048$ & & \\
2695.4 & A$^2$ & P & $0.497 \pm 0.028$ & & \\
2695.4 & A & S$^b$ & $0.507 \pm 0.029$ & $0.97 \pm 0.12$ & $4.45 \pm 0.18$\\
2963.6 & A & S & $0.484 \pm 0.028$ & $0.96 \pm 0.12$ & $3.99 \pm 0.16$\\
2963.6 & A & P & $0.479 \pm 0.027$ & & \\
3047.3 & A & P & $0.473 \pm 0.027$ & & \\
3047.3 & A & S & $0.479 \pm 0.027$ & $0.96 \pm 0.12$ & $3.76 \pm 0.15$\\
3049.3 & A & P & $0.481 \pm 0.027$ & & \\
3049.3 & A & S & $0.484 \pm 0.028$ & $1.01 \pm 0.13$ & $4.29 \pm 0.17$\\
3388.4 & A & P & $0.495 \pm 0.028$ & & \\
3388.4 & A & S & $0.499 \pm 0.028$ & $1.02 \pm 0.13$ & $4.27 \pm 0.17$\\
3404.0 & A & P & $0.490 \pm 0.028$ & & \\
3474.0 & A & P & $0.491 \pm 0.028$ & & \\
3475.3 & A & S & $0.496 \pm 0.028$ & $1.00 \pm 0.12$ & $4.03 \pm 0.16$\\
3714.0 & C$^3$ & P & $0.462 \pm 0.015$ & & \\
3745.0 & C & P & $0.470 \pm 0.015$ & & \\
3769.0 & C & P & $0.465 \pm 0.015$ & & \\
3772.0 & A & P & $0.476 \pm 0.027$ & & \\
3797.4 & A & S & $0.469 \pm 0.027$ & $0.90 \pm 0.11$ & $4.13 \pm 0.17$\\
3798.0 & A & P & $0.473 \pm 0.027$ & & \\
4050.0 & C & P & $0.474 \pm 0.012$ & & \\
4058.0 & C & P & $0.474 \pm 0.015$ & & \\
4066.0 & C & P & $0.476 \pm 0.015$ & & \\
4068.7 & A & S & $0.485 \pm 0.028$ & $0.94 \pm 0.12$ & $4.46 \pm 0.18$\\
4068.7 & A & P & $0.483 \pm 0.028$ & & \\
4085.0 & C & P & $0.471 \pm 0.015$ & & \\
4091.4 & A & S & $0.475 \pm 0.027$ & $0.93 \pm 0.12$ & $4.37 \pm 0.18$\\
4092.0 & A & P & $0.477 \pm 0.027$ & & \\
4127.0 & C & P & $0.474 \pm 0.015$ & & \\
4145.3 & A & P & $0.477 \pm 0.027$ & & \\
4145.3 & A & S & $0.473 \pm 0.027$ & $0.94 \pm 0.12$ & $3.85 \pm 0.16$\\
4201.3 & A & P & $0.474 \pm 0.027$ & & \\
4201.3 & A & S & $0.460 \pm 0.026$ & $0.88 \pm 0.11$ & $4.13 \pm 0.17$\\
4204.0 & C & P & $0.467 \pm 0.015$ & & \\
\multicolumn{6}{l}{ }\\
\hline
\end{tabular*}
\end{table*}

\begin{table*}[htbp]
\centering
\begin{tabular*}{\textwidth}{c @{\extracolsep{\fill}} cccccc}
\multicolumn{6}{l}{{\footnotesize {\bfseries Table 2.} Continued.}}\\
\multicolumn{6}{l}{ }\\
\hline
\multicolumn{6}{l}{ }\\
4238.0 & C & P & $0.499 \pm 0.019$ & & \\
4435.7 & A & S & $0.491 \pm 0.028$ & $0.97 \pm 0.12$ & $4.66 \pm 0.19$\\
4435.7 & A & P & $0.496 \pm 0.028$ & & \\
4472.5 & A & P & $0.483 \pm 0.028$ & & \\
4472.5 & A & S & $0.482 \pm 0.027$ & $0.92 \pm 0.11$ & $4.38 \pm 0.18$\\
4513.4 & A & P & $0.466 \pm 0.027$ & & \\
4513.4 & A & S & $0.469 \pm 0.027$ & $1.03 \pm 0.13$ & $4.03 \pm 0.16$\\
4747.0 & C & P & $0.479 \pm 0.018$ & & \\
4769.0 & C & P & $0.481 \pm 0.015$ & & \\
4806.0 & C & P & $0.499 \pm 0.016$ & & \\
4807.6 & A & S & $0.495 \pm 0.028$ & $0.97 \pm 0.12$ & $4.23 \pm 0.17$\\
4807.6 & A & P & $0.498 \pm 0.028$ & & \\
4835.4 & A & S & $0.497 \pm 0.028$ & $0.92 \pm 0.11$ & $4.66 \pm 0.19$\\
4836.0 & A & P & $0.486 \pm 0.028$ & & \\
4850.0 & C & P & $0.412 \pm 0.029$ & & \\
4852.0 & C & P & $0.461 \pm 0.017$ & & \\
4862.0 & C & P & $0.478 \pm 0.015$ & & \\
4884.3 & A & P & $0.488 \pm 0.028$ & & \\
4884.3 & A & S & $0.492 \pm 0.028$ & $0.95 \pm 0.12$ & $3.92 \pm 0.16$\\
4886.0 & C & P & $0.466 \pm 0.017$ & & \\
4892.0 & C & P & $0.468 \pm 0.015$ & & \\
4913.0 & C & P & $0.470 \pm 0.015$ & & \\
4914.4 & A & P & $0.485 \pm 0.028$ & & \\
4914.4 & A & S & $0.485 \pm 0.028$ & $1.04 \pm 0.13$ & $4.40 \pm 0.18$\\
4945.0 & C & P & $0.469 \pm 0.015$ & & \\
4979.0 & C & P & $0.455 \pm 0.017$ & & \\
5133.0 & C & P & $0.459 \pm 0.015$ & & \\
5155.0 & C & P & $0.454 \pm 0.015$ & & \\
5183.0 & C & P & $0.453 \pm 0.014$ & & \\
5247.0 & C & P & $0.437 \pm 0.014$ & & \\
5273.0 & C & P & $0.416 \pm 0.013$ & & \\
5298.0 & C & P & $0.421 \pm 0.016$ & & \\
5336.0 & C & P & $0.406 \pm 0.013$ & & \\
5511.0 & C & P & $0.378 \pm 0.015$ & & \\
5542.0 & C & P & $0.377 \pm 0.017$ & & \\
5565.0 & C & P & $0.375 \pm 0.012$ & & \\
5573.0 & C & P & $0.372 \pm 0.013$ & & \\
5591.0 & C & P & $0.374 \pm 0.012$ & & \\
5602.0 & C & P & $0.363 \pm 0.012$ & & \\
5618.0 & C & P & $0.361 \pm 0.014$ & & \\
5636.0 & C & P & $0.360 \pm 0.013$ & & \\
5653.3 & A & P & $0.345 \pm 0.020$ & & \\
5653.3 & A & S & $0.350 \pm 0.020$ & $1.35 \pm 0.17$ & $4.81 \pm 0.19$\\
5694.0 & C & P & $0.357 \pm 0.013$ & & \\
5894.5 & A & P & $0.374 \pm 0.021$ & & \\
5894.5 & A & S & $0.347 \pm 0.020$ & $1.34 \pm 0.17$ & $5.41 \pm 0.22$\\
5915.4 & A & S & $0.358 \pm 0.020$ & $1.28 \pm 0.16$ & $4.77 \pm 0.19$\\
5916.0 & A & P & $0.352 \pm 0.020$ & & \\
5923.0 & C & P & $0.361 \pm 0.013$ & & \\
5939.0 & C & P & $0.361 \pm 0.012$ & & \\
5945.0 & C & P & $0.350 \pm 0.013$ & & \\
5950.0 & C & P & $0.348 \pm 0.020$ & & \\
5956.0 & C & P & $0.349 \pm 0.013$ & & \\
5979.0 & C & P & $0.355 \pm 0.013$ & & \\
5984.0 & C & P & $0.352 \pm 0.013$ & & \\
5985.3 & A & P & $0.356 \pm 0.020$ & & \\
5985.3 & A & S & $0.347 \pm 0.020$ & $1.08 \pm 0.14$ & $5.10 \pm 0.21$\\
6008.0 & C & P & $0.357 \pm 0.013$ & & \\
6039.0 & C & P & $0.362 \pm 0.014$ & & \\
6238.4 & A & S & $0.377 \pm 0.021$ & $1.08 \pm 0.14$ & $4.52 \pm 0.18$\\
\multicolumn{6}{l}{ }\\
\hline
\end{tabular*}
\end{table*}

\begin{table*}[htbp]
\centering
\begin{tabular*}{\textwidth}{c @{\extracolsep{\fill}} cccccc}
\multicolumn{6}{l}{{\footnotesize {\bfseries Table 2.} Continued.}}\\
\multicolumn{6}{l}{ }\\
\hline
\multicolumn{6}{l}{ }\\
6238.4 & A & P & $0.386 \pm 0.022$ & & \\
6255.0 & C & P & $0.394 \pm 0.013$ & & \\
6272.0 & C & P & $0.392 \pm 0.012$ & & \\
6272.0 & L$^4$ & S & $0.396 \pm 0.023$ & $1.18 \pm 0.15$ & $4.94 \pm 0.20$\\
6299.0 & C & P & $0.387 \pm 0.012$ & & \\
6307.0 & C & P & $0.392 \pm 0.013$ & & \\
6314.0 & C & P & $0.391 \pm 0.012$ & & \\
6325.0 & C & P & $0.395 \pm 0.015$ & & \\
6332.0 & C & P & $0.400 \pm 0.013$ & & \\
6338.0 & C & P & $0.396 \pm 0.015$ & & \\
6340.3 & L & S & $0.403 \pm 0.023$ & $1.36 \pm 0.17$ & $5.50 \pm 0.22$\\
6358.0 & C & P & $0.400 \pm 0.013$ & & \\
6374.0 & C & P & $0.401 \pm 0.014$ & & \\
6396.3 & L & S & $0.413 \pm 0.024$ & $1.27 \pm 0.16$ & $5.44 \pm 0.22$\\
6400.0 & C & P & $0.405 \pm 0.013$ & & \\
6421.4 & L & S & $0.425 \pm 0.024$ & $1.31 \pm 0.16$ & $5.20 \pm 0.21$\\
6429.0 & C & P & $0.408 \pm 0.013$ & & \\
6568.0 & C & P & $0.405 \pm 0.013$ & & \\
6588.0 & C & P & $0.410 \pm 0.013$ & & \\
6749.0 & L & S & $0.402 \pm 0.023$ & $1.02 \pm 0.13$ & $4.83 \pm 0.20$\\
6786.3 & L & S & $0.428 \pm 0.024$ & $1.01 \pm 0.13$ & $4.24 \pm 0.17$\\
7093.3 & L & S & $0.466 \pm 0.027$ & $1.13 \pm 0.14$ & $5.15 \pm 0.21$\\
\multicolumn{6}{l}{ }\\
\hline
\multicolumn{6}{l}{{\footnotesize $^1$0.5 m telescope at Climenhaga Observatory, University of Victoria (USA; \citealt{Lew99})}}\\
\multicolumn{6}{l}{{\footnotesize $^2$1.82 m Copernico telescope at Asiago Observatory (Italy; \citealt{Tre07})}}\\
\multicolumn{6}{l}{{\footnotesize $^3$Catalina Sky Survey \citep{Dra09}}}\\
\multicolumn{6}{l}{{\footnotesize $^4$1.52 m telescope at Loiano Observatory (Italy; \citealt{Tre14})}}\\
\multicolumn{6}{l}{{\footnotesize $^a$Photometry}}\\
\multicolumn{6}{l}{{\footnotesize $^b$Spectroscopy}}\\
\end{tabular*}
\end{table*}

\end{document}